\documentclass[10pt,journal,compsoc]{IEEEtran}
\usepackage{stmaryrd}

\usepackage{amsmath}
\usepackage{amsthm}
\usepackage{amsfonts}
\usepackage{txfonts}
\usepackage{algorithm}
\usepackage{algorithmic}
\usepackage{enumerate}
\usepackage{graphicx}
\newtheorem{Thm}{Theorem}
\newtheorem{Lem}[Thm]{Lemma}
\newtheorem{Def}{Definition}
\newtheorem{Exm}{Example}

\newtheorem{Cor}[Thm]{Corollary}

\ifCLASSOPTIONcompsoc
  \usepackage[nocompress]{cite}
\else
  \usepackage{cite}
\fi
\ifCLASSINFOpdf
\else
\fi
\begin{document}
\bstctlcite{BSTcontrol}
%
\title{An Improved Decision Procedure for Linear Time Mu-Calculus}
%
%
%
%

\author{Yao~Liu,
        Zhenhua~Duan
        and~Cong~Tian
\IEEEcompsocitemizethanks{\IEEEcompsocthanksitem The authors are with the ICTT and ISN Laboratory, Xidian University, Xi'an 710071, China.
E-mail: yao\_liu@stu.xidian.edu.cn, \{zhhduan,ctian\}@mail.xidian.edu.cn.}
}

\IEEEtitleabstractindextext{%
\begin{abstract}
An improved Present Future form (PF form) for linear time $\mu$-calculus ($\nu$TL) is presented in this paper. In particular, the future part of the new version turns into the conjunction of elements in the closure of a formula. We show that every closed $\nu$TL formula can be transformed into the new PF form. Additionally, based on the PF form, an algorithm for constructing Present Future form Graph (PFG), which can be utilized to describe models of a formula, is given. Further, an intuitive and efficient decision procedure for checking satisfiability of the guarded fragment of $\nu$TL formulas based on PFG is proposed and implemented in C++. The new decision procedure has the best time complexity over the existing ones despite the cost of exponential space. Finally, a PFG-based model checking approach for $\nu$TL is discussed where a counterexample can be obtained visually when a model violates a property.
\end{abstract}

}

\maketitle

\IEEEdisplaynontitleabstractindextext

%
\IEEEpeerreviewmaketitle

\IEEEraisesectionheading{\section{Introduction}\label{sec:introduction}}

%
%
%
%
\IEEEPARstart{L}{inear} time $\mu$-calculus ($\nu$TL) \cite{barringer1986really}, linear time counterpart of modal $\mu$-calculus \cite{Koz83}, extends LTL \cite{Pnu77} with least and greatest fixpoint operators. It is a formalism succinct in syntax and strong in expressive power which captures the expression of full $\omega$-regular properties \cite{EC80,barringer1985compositional}. Hence, it is useful for specifying and verifying various properties of concurrent programs and has received ever growing interest in the past few decades. From an application point of view, it is of great significance to establish a decision procedure for checking satisfiability of $\nu$TL formulas. The work, however, is not easy due to the nesting of fixpoint operators.

Satisfiability and model checking \cite{clarke1986automatic} are two main decision problems for $\nu$TL, which are both PSPACE-complete in complexity \cite{vardi1988temporal}. By satisfiability we denote the problem to find a decision procedure for determining whether a formula is satisfiable, while by model checking we mean the problem to decide whether all paths of a given Kripke structure satisfy a certain property. Moreover, decision procedures for checking satisfiability always play a critical role in deriving model checking approaches.

A lot of work has been done for achieving efficient decision procedures. The major milestone of the decision problems for modal $\mu$-calculus is made by Streett and Emerson \cite{streett1989automata} who introduce the notion of well-founded pre-models and apply automata theory to check satisfiability. Related methods \cite{EJS93,KVW00} translate a formula into an equivalent alternating tree automaton and then check for emptiness. In \cite{vardi1988temporal}, Vardi first adapts Streett and Emerson's method to $\nu$TL with past operators which yields an algorithm running in $2^{O(|\phi|^4)}$. Later, Banieqbal and Barringer \cite{Banieqbal1987} show that if a formula has a model, then it is able to generate a good Hintikka structure which can be further transformed into a good path searching problem from a graph. Their algorithm is equivalent in time complexity to Vardi's but runs in exponential space. In \cite{stirling1990ccs}, Stirling and Walker present a tableau characterisation for $\nu$TL's decision problems without mentioning complexity issues. Bradfield, Esparza and Mader \cite{bradfield1996effective} improve the system of Stirling and Walker based on the work in \cite{kaivola1995simple} by simplifying the success conditions for a tableau and their algorithm runs in $2^{O(|\phi|^2\log|\phi|)}$. In \cite{dhl06}, Dax, Hofmann and Lange propose a decision procedure for checking validity of $\nu$TL formulas running in $2^{O(|\phi|^2\log|\phi|)}$ and implemented in OCAML. To the best of our knowledge, all these existing decision methods mentioned above except for \cite{dhl06} are relatively complicated and concerned with theoretical aspects rather than practical applications. Therefore, we are motivated to formalize a more efficient and practical decision procedure.

To this end, a new Present Future form (PF form) for $\nu$TL formulas is presented in this paper and we prove that every closed $\nu$TL formula can be transformed into this form. Compared with \cite{PFForm14}, the definition of the new PF form, which still consists of the present and future parts, is more elegant: the present part remains unchanged while the future part turns into the conjunction of elements in the closure of a given formula. This further facilitates the proof of finiteness of Present Future form Graph (PFG) which can be used to describe models of a formula. A path in a PFG characterizes exactly a pre-model \cite{vardi1988temporal,streett1989automata} of the corresponding formula. Additionally, an algorithm, based on PF form, for constructing PFG is given. In a PFG, an edge may be associated with a mark which is a subset of variables occurring in the formula and utilized to keep track of the infinite unfolding problem for least fixpoint formulas. Further, a decision procedure for checking satisfiability of the guarded fragment of $\nu$TL formulas based on PFG is presented. It is realized, with the help of marks, by searching for a $\nu$-path in a PFG on which no least fixpoint formula unfolds itself infinitely. Moreover, the decision procedure has been implemented in C++. The result shows that our method improves the current best time complexity, $2^{O(|\phi|^2\log|\phi|)}$ \cite{bradfield1996effective,kaivola1995simple,dhl06}, to $2^{O(|\phi|)}$ despite the cost of exponential space.

According to the proposed decision procedure, a PFG-based model checking approach for $\nu$TL is also proposed. To do so, first, an algorithm for constructing the product of a Kripke structure and a PFG is presented. Subsequently, we apply the notion of $\nu$-paths in PFGs to the product graphs. Further, given a Kripke structure $M$ and a desired property $\phi$, the model checking approach is achieved by searching for a $\nu$-path in the product graph of $M$ and the PFG of $\neg\phi$. If such a path can be found, we will obtain a counterexample; otherwise, $M$ satisfies $\phi$.

The idea of this paper is inspired by the normal form and normal form
graph of Propositional Projection Temporal Logic (PPTL) \cite{duan1996extended,duan2006book} which have played a vital role in obtaining a decision procedure for
checking the satisfiability
\cite{DT2007Decidability,DT2010ImprovedDecision,DTpracticaldecision14}. Compared with the existing methods for checking satisfiability of $\nu$TL formulas, our decision procedure has the following advantages: (1) it does not depend on automata theory by considering PFGs; (2) it is more efficient in time and practical meanwhile; (3) it gives good insight into why and how a given formula is satisfiable through its PFG; (4) when a Kripke structure violates a property, it intuitively reflects that why a path is a counterexample through the corresponding product graph.

To summarize, our contributions are as follows:
\begin{itemize}
  \item We define a new PF form for $\nu$TL formulas and prove that every closed $\nu$TL formula can be transformed into this form.
  \item We provide an algorithm for constructing PFG which can be used to describe models of a formula. During the constructing process, marks are technically added, which are useful in keeping track of the infinite unfolding problem for least fixpoint formulas.
  \item We introduce the notion of $\nu$-paths and present a decision procedure for checking satisfiability of the guarded fragment of $\nu$TL formulas by finding a $\nu$-path in a PFG.
  \item We show that our decision procedure has the current best time complexity. We implement the decision procedure in C++ and experimental results show that our algorithm performs better than the one given in \cite{dhl06}.
  \item We apply the notion of $\nu$-paths in PFGs to the product graphs and propose a PFG-based model checking approach for $\nu$TL.
\end{itemize}

The rest of this paper is organized as follows. The syntax and semantics of $\nu$TL and some basic notions are introduced in Section \ref{sectionPreliminaries}. The new PF form of $\nu$TL formulas is presented in Section \ref{sectionPFForm}. Section \ref{sectionPFGConstruction} describes an algorithm for constructing PFG and the decision procedure for checking satisfiability of the guarded fragment of $\nu$TL formulas based on PFG is given in Section \ref{sectionDecisionProcedure}. Section \ref{sectionModelChecking} presents a model checking approach for $\nu$TL based on PFG. Related work is discussed in section \ref{sectionRelatedWork}. Conclusions are drawn in Section \ref{sectionConclusion}.

\section{Preliminaries}\label{sectionPreliminaries}
\subsection{Syntax and Semantics of $\nu$TL}
Let $\mathcal {P}$ be a set of atomic propositions, and $\mathcal
{V}$ a set of variables. $\nu$TL formulas are constructed based on the following syntax:
\[\phi :: = p~|~\neg p~|~X~|~\phi \vee \phi~|~\phi \wedge \phi~|\bigcirc \phi~|~\mu X.\phi~|~\nu X.\phi\] where $p$ ranges over $\mathcal {P}$ and $X$ over $\mathcal
{V}$.

We use $\sigma$ to denote either $\mu$ or $\nu$. An occurrence of a variable $X$ in a formula is called \emph{free} when it does not lie within the scope of $\sigma X$; it is called \emph{bound} otherwise. A formula is called \emph{closed} when it contains no free variables. Given two $\nu$TL formulas $\phi_1$ and $\phi_2$, we say $\phi_1\preceq\phi_2$ iff $\phi_2$ is a subformula of $\phi_1$, and $\phi_1\prec\phi_2$ iff $\phi_2$ is a proper subformula of $\phi_1$. We write $\phi[\phi'/Y]$ for the result of simultaneously substituting $\phi'$ for all free occurrences of variable $Y$ in $\phi$. For each variable $X$ in a formula, we assume that $X$ is bound at most once. Thus, it can be seen that all formulas constructed by the syntax above are in positive normal form \cite{stirling1991modal}, i.e. negations can be applied only to atomic propositions and each variable occurring in a formula is bound at most once.

For each bound variable $X$ in formula $\phi$, the unique subformula of $\phi$ in the form of $\sigma X.\varphi$ is said to be \emph{identified by} $X$. The bound variables in $\phi$ can be partially ordered based on the nesting of their identified fixpoint formulas. Specifically, given two bound variables $X$ and $Y$ in $\phi$, we say $X$ is \emph{higher than} $Y$ iff the fixpoint formula identified by $Y$ is a proper subformula of the one identified by $X$.

A formula is called a \emph{guarded} one if, for each bound variable $X$ in that formula, every occurrence of $X$ is in the scope of a $\bigcirc$ operator. Every formula can be transformed into an equivalent one in guarded form \cite{walukiewicz2000completeness}. Note that the transformation causes an exponential increase in the size of a formula in the worst case \cite{bruse2015guarded}.
\begin{Exm}
Translating formula $\nu X.(p\wedge \mu Y.(q\vee X\wedge\bigcirc Y))$ into guarded form.
\end{Exm}
\[
\begin{array}{ll}
\, & \nu X.(p\wedge \mu Y.(q\vee X\wedge\bigcirc Y)) \\
\equiv & \nu X.(p\wedge (q\vee X\wedge\bigcirc \mu Y.(q\vee X\wedge\bigcirc Y))) \\
\, & by\;law\;\sigma X.\phi \equiv \phi[\sigma X.\phi/X] \\
\equiv & \nu X.(p\wedge q \vee p\wedge\bigcirc \mu Y.(q\vee X\wedge\bigcirc Y)) \\
\, & by\;law\;\nu X.(X\wedge\phi \vee \varphi) \equiv \nu X.(\phi \vee \varphi)
\end{array}
\]

$\nu$TL formulas are interpreted over linear time structures. A \textit{linear time structure} over $\mathcal {P}$ is a function $\mathcal {K}$: $\mathbb {N}\rightarrow 2^{\mathcal {P}}$ where $\mathbb{N}$ denotes the set of natural numbers. The semantics of a $\nu$TL formula $\phi$, relative to $\mathcal {K}$ and an environment $e:\mathcal {V}\rightarrow 2^{\mathbb{N}}$, is inductively defined as follows:
\[
\begin{array}{rcl}
\llbracket p\rrbracket^{\mathcal {K}}_e & \coloneqq & \{i\in\mathbb{N}\;|\;p\in \mathcal {K}(i)\} \\
\llbracket \neg p\rrbracket^{\mathcal {K}}_e & \coloneqq & \{i\in\mathbb{N}\;|\;p\notin \mathcal {K}(i)\} \\
\llbracket X\rrbracket^{\mathcal {K}}_e & \coloneqq & e(X)\\
\llbracket \varphi\vee\psi\rrbracket^{\mathcal {K}}_e & \coloneqq & \llbracket \varphi\rrbracket^{\mathcal {K}}_e \cup \llbracket \psi\rrbracket^{\mathcal {K}}_e\\
\llbracket \varphi\wedge\psi\rrbracket^{\mathcal {K}}_e & \coloneqq & \llbracket \varphi\rrbracket^{\mathcal {K}}_e \cap \llbracket \psi\rrbracket^{\mathcal {K}}_e\\
\llbracket \bigcirc\varphi\rrbracket^{\mathcal {K}}_e & \coloneqq & \{i\in\mathbb{N}\;|\;i+1\in \llbracket \varphi\rrbracket^{\mathcal {K}}_e\}\\
\llbracket \mu X.\varphi\rrbracket^{\mathcal {K}}_e & \coloneqq & \bigcap\{W\subseteq \mathbb{N}\;|\;\llbracket\varphi\rrbracket^{\mathcal {K}}_{e[X\mapsto W]}\subseteq W\}\\
\llbracket \nu X.\varphi\rrbracket^{\mathcal {K}}_e & \coloneqq & \bigcup\{W\subseteq \mathbb{N}\;|\;W\subseteq \llbracket\varphi\rrbracket^{\mathcal {K}}_{e[X\mapsto W]}\}
\end{array}
\]
where $e[X\mapsto W]$ is the environment $e'$ agreeing with $e$ except for $e'(X)=W$. $e$ is used to evaluate free variables and can be dropped when $\phi$ is closed.

For a given formula $\phi$, we say $\phi$ is true at state $i$ of linear time structure $\mathcal {K}$, denoted by $\mathcal {K},i\models\phi$, iff $i\in \llbracket\phi\rrbracket^{\mathcal {K}}_e$. We say $\phi$ is \emph{valid}, denoted by $\models\phi$, iff $\mathcal {K},j\models\phi$ for all linear time structures $\mathcal {K}$ and all states $j$ of $\mathcal {K}$; $\phi$ is \emph{satisfiable} iff there exists a linear time structure $\mathcal {K}$ and a state $j$ of $\mathcal {K}$ such that $\mathcal {K},j\models\phi$.

\subsection{Approximant}
Let $Ord$ denote the class of ordinals. \textit{Approximants} of fixpoint formulas are defined inductively by: $\mu^{0}X.\phi = \bot$, $\nu^{0}X.\phi = \top$, $\sigma^{\alpha+1}X.\phi = \phi[\sigma^{\alpha}X.\phi/X]$, $\mu^{\lambda}X.\phi = \bigvee_{\alpha<\lambda} \mu^{\alpha}X.\phi$ and $\nu^{\lambda}X.\phi = \bigwedge_{\alpha<\lambda}\nu^{\alpha}X.\phi$ where $\alpha,\lambda\in Ord$. In particular, $\lambda$ is a limit ordinal.

The following lemma \cite{tarski1955lattice} is a standard result about approximants.
\begin{Lem}\label{ApproximantLem}
For a linear time structure $\mathcal{K}$, we say $\mathcal{K},0\models \nu X.\phi$ iff $\forall \alpha \in Ord$, $\mathcal{K},0\models \nu^{\alpha} X.\phi$ and $\mathcal{K},0\models \mu Y.\phi$ iff $\exists \alpha \in Ord$, $\mathcal{K},0\models \mu^{\alpha} Y.\phi$. \emph{\textbf{(\cite{tarski1955lattice})}}
\end{Lem}
Note that in both cases above, $\alpha$ is not a limit ordinal.

Let $\phi$ be a closed $\nu$TL formula with exactly $n$ $\mu$-variables: $X_1,\ldots,X_n$ such that $X_i$ is higher than $X_j$ implies $i<j$. A $\mu$-$signature$ for $\phi$ is a tuple $\zeta=(\alpha_1,\ldots,\alpha_n)\in (\mathbb{N}\cup \{\omega\})^n$ where each $\alpha_i$ is an ordinal. A $\mu$-signature \emph{with respect to} a variable $Y$ in $\phi$ is the prefix $(\alpha_1,\ldots,\alpha_i)$ of $\zeta$ such that $Y=X_i$ when $Y$ is a $\mu$-variable, or $X_i$ is the last $\mu$-variable higher than $Y$ when $Y$ is a $\nu$-variable. We write $\zeta(i)$ for the $i$-th component of $\zeta$. For two $\mu$-signatures $\zeta_1$ and $\zeta_2$ for $\phi$, we write $\zeta_1<\zeta_2$ to mean that $\zeta_1$ lexicographically precedes $\zeta_2$, i.e. $\zeta_1(j)<\zeta_2(j)$ and $\zeta_1(i)=\zeta_2(i)$ for some $j$ and each $i<j$. Note that the lexicographic ordering on $\mu$-signatures is well-founded.

For a linear time structure $\mathcal {K}$ and a state $j$ of $\mathcal {K}$, we say $\mathcal {K},j\models_{\zeta}\phi$ if $(\mathcal {K},j)$ is a model of $\phi$ resulting from $\phi$ with every least fixpoint subformula $\mu X_i.\phi_i$ of $\phi$ being interpreted by $\mu^{\zeta(i)}X_i.\phi_i$.
\subsection{Closure}
The \textit{closure} $CL(\phi)$ of a formula $\phi$, based on \cite{FL79}, is the least set of formulas such that
\begin{enumerate}[(i)]
  \item $\phi,true \in CL(\phi),$
  \item if $\varphi\vee\psi$ or $\varphi\wedge\psi\in CL(\phi)$, then $\varphi \in CL(\phi)$ and $\psi \in CL(\phi),$
  \item if $\bigcirc\varphi \in CL(\phi)$, then $\varphi \in CL(\phi),$
  \item if $\sigma X.\varphi \in CL(\phi)$, then $\varphi[\sigma X.\varphi/X] \in CL(\phi).$
\end{enumerate}

\begin{Exm}
The closure of formula $\nu X.\mu Y.(\bigcirc Y \vee p\wedge\bigcirc X)$.
\end{Exm}
$CL(\nu X.\mu Y.(\bigcirc Y \vee p\wedge\bigcirc X))=$\\
\centerline{$\{\nu X.\mu Y.(\bigcirc Y \vee p\wedge\bigcirc X),true,$}
\centerline{$\mu Y.(\bigcirc Y \vee p\wedge\bigcirc \nu X.\mu Y.(\bigcirc Y \vee p\wedge\bigcirc X)),$} \\
\centerline{$\bigcirc \mu Y.(\bigcirc Y \vee p\wedge\bigcirc \nu X.\mu Y.(\bigcirc Y \vee p\wedge\bigcirc X))$}\\
\centerline{$\vee p\wedge\bigcirc \nu X.\mu Y.(\bigcirc Y \vee p\wedge\bigcirc X),
$}\\
\centerline{$\bigcirc \mu Y.(\bigcirc Y \vee p\wedge\bigcirc \nu X.\mu Y.(\bigcirc Y \vee p\wedge\bigcirc X)),
$}
\centerline{$p\wedge\bigcirc \nu X.\mu Y.(\bigcirc Y \vee p\wedge\bigcirc X),
$}
\centerline{$p,\;\bigcirc \nu X.\mu Y.(\bigcirc Y \vee p\wedge\bigcirc X)\}
$}

It has been proved that the size of $CL(\phi)$ is linear in the size of $\phi$ (denoted by $|\phi|$) \cite{FL79}.
\subsection{Dependency Relationship}
\begin{Def}\label{DependencyDef}
For two formulas $\sigma X.\phi$ and $\sigma Y.\phi'$ where $\sigma X.\phi \prec \sigma Y.\phi'$, we say $Y$ \emph{depends on} $X$, denoted by $X\lhd Y$, iff $X$ occurs free in $\phi'$.
\end{Def}

Note that the dependency relationship is transitive in a formula.
\begin{Exm}\label{AlterVariableExm}
Dependency relationship between variables.
\begin{itemize}
\item[I.] $\nu X. (\bigcirc X \wedge \mu Y.(p\vee \bigcirc Y))$
\item[II.] $\nu X. \mu Y.(\bigcirc Y\vee p\wedge \bigcirc X)\vee \mu Z. \nu W.(\bigcirc Z\vee q\wedge \bigcirc W)$
\item[III.] $\mu X. \nu Y.(\bigcirc X \vee \mu Z.\bigcirc(Z\vee Y\wedge p))$
\end{itemize}
\end{Exm}

In formula I, $X$ and $Y$ do not depend on each other. In formula II, we have $X\lhd Y$ and $Z\lhd W$, while in formula III we have $X\lhd Y\lhd Z$.
\section{PF Form of $\nu$TL Formulas}\label{sectionPFForm}
In this section, we first define PF form of $\nu$TL formulas and then prove that every closed $\nu$TL formula can be transformed into this form.
\subsection{PF Form}
\begin{Def}\label{PFFormDef}
Let $\phi$ be a closed $\nu$TL formula, $\mathcal {P}_{\phi}$ the set of atomic propositions appearing in $\phi$. \emph{PF form} of $\phi$ is defined by:
\[\phi\equiv
\bigvee_{i=1}^{n}(\phi_{p_i}\wedge\bigcirc \phi_{f_i})\]
where $\phi_{p_i}\equiv\bigwedge_{h=1}^{n_1}\dot{p}_{ih}$, $p_{ih} \in \mathcal {P}_{\phi}$ for each h ($\dot{r}$ denotes either $r$ or $\neg r$ for each $r \in \mathcal {P}_{\phi}$) and $\phi_{f_i}\equiv\bigwedge_{m=1}^{n_2}\phi_{im}$, $\phi_{im} \in CL(\phi)$ for each m.
\end{Def}

The main difference between the PF form presented here and the one in \cite{PFForm14} lies in the future part: in this paper, the future part is the conjunction of elements in the closure of a given formula rather than a closed formula in \cite{PFForm14}. Thus, it can be seen that the PF form presented here is more rigorous in structure and this will dramatically simplify the proof of finiteness of PFG.

In the following, we prove that every closed $\nu$TL formula can be transformed into PF form. For technical reasons, from now on we confine ourselves only to guarded formulas with no $\vee$ appearing as the main operator under each $\bigcirc$ operator. This can be easily achieved by pushing $\bigcirc$ operators inwards using the equivalence $\bigcirc(\phi_1\vee\phi_2)\equiv\bigcirc\phi_1\vee\bigcirc\phi_2$.
\begin{Thm}\label{PFFormThm}
Every closed $\nu$TL formula $\varphi$ can be transformed into PF form.
\end{Thm}
\textit{Proof.} Let $Conj(\psi)$ represent the set of all conjuncts in formula $\psi$. The proof proceeds by induction on the structure of $\varphi$.
\begin{itemize}
\item[$\bullet$] \textbf{Base case:}
\item[--] $\varphi$ is $p$ (or $\neg p$): $p$ (or $\neg p$) can be transformed as:
\[p\equiv p\wedge \bigcirc true\;(or\;\neg p\equiv \neg p\wedge \bigcirc true)\]
The theorem holds obviously in these two cases.

\item[$\bullet$] \textbf{Induction:}
\item[--] $\varphi$ is $\bigcirc\phi$: $\bigcirc\phi$ can be written as:
\[\bigcirc\phi \equiv \bigvee_{i} (true \wedge \bigcirc \phi_i)\]
For each $\phi_c \in Conj(\phi_i)$, we have $\phi_c \in CL(\varphi)$ since $\phi \in CL(\varphi)$. Hence, $\varphi$ can be transformed into PF form in this case.
\item[--] $\varphi$ is $\phi_1 \vee \phi_2$: by induction hypothesis, both $\phi_1$ and $\phi_2$ can be transformed into PF form:
\[\phi_1\equiv\bigvee_{i=1}^{n}(\phi_{1p_i}\wedge\bigcirc
\phi_{1f_i}),\;\phi_2\equiv\bigvee_{j=1}^{m}(\phi_{2p_j}\wedge\bigcirc
\phi_{2f_j})\]
    where $\phi_{1c} \in Conj(\phi_{1f_i})$ and $\phi_{1c} \in CL(\phi_1)$, $\phi_{2c} \in Conj(\phi_{2f_j})$ and $\phi_{2c} \in CL(\phi_2)$, for each $i$ and $j$. Then, we have
    \[\varphi\equiv\phi_1 \vee \phi_2\equiv\bigvee_{i=1}^{n}(\phi_{1p_i}\wedge\bigcirc
\phi_{1f_i}) \vee \bigvee_{j=1}^{m}(\phi_{2p_j}\wedge\bigcirc
\phi_{2f_j})\]
Since $\phi_1 \vee \phi_2 \in CL(\varphi)$, we have $\phi_1, \phi_2 \in CL(\varphi)$. For each $\phi_{1c} \in Conj(\phi_{1f_i})$, by induction hypothesis, we have $\phi_{1c} \in CL(\phi_1)$. Therefore, $\phi_{1c} \in CL(\varphi)$. Similarly, we can obtain that each $\phi_{2c}\in CL(\varphi)$. Thus, $\varphi$ can be transformed into PF form in this case.

\item[--] $\varphi$ is $\phi_1 \wedge \phi_2$: by induction hypothesis, both $\phi_1$ and $\phi_2$ can be transformed into PF form: \[\phi_1\equiv\bigvee_{i=1}^{n}(\phi_{1p_i}\wedge\bigcirc
\phi_{1f_i}),\;\phi_2\equiv\bigvee_{j=1}^{m}(\phi_{2p_j}\wedge\bigcirc
\phi_{2f_j})\]
where $\phi_{1c} \in Conj(\phi_{1f_i})$ and $\phi_{1c} \in CL(\phi_1)$, $\phi_{2c} \in Conj(\phi_{2f_j})$ and $\phi_{2c} \in CL(\phi_2)$, for each $i$ and $j$. Then $\varphi$ can be further converted into:
\[\varphi\equiv\phi_1 \wedge
\phi_2\equiv(\bigvee_{i=1}^{n}(\phi_{1p_i}\wedge\bigcirc
\phi_{1f_i})) \wedge (\bigvee_{j=1}^{m}(\phi_{2p_j}\wedge\bigcirc
\phi_{2f_j}))\] \[~~~~~~~~~~~\equiv \bigvee_{i=1}^{n}\bigvee_{j=1}^{m}(\phi_{1p_i} \wedge \phi_{2p_j} \wedge \bigcirc (\phi_{1f_i} \wedge \phi_{2f_j}))\]
Since $\phi_1 \wedge \phi_2 \in CL(\varphi)$, we have $\phi_1, \phi_2 \in CL(\varphi)$. For each $\phi_{1c} \in Conj(\phi_{1f_i})$, by induction hypothesis, we have $\phi_{1c} \in CL(\phi_1)$. Hence, $\phi_{1c} \in CL(\varphi)$. Similarly, we can obtain that each $\phi_{2c}\in CL(\varphi)$. Therefore, all conjuncts behind $\bigcirc$ operators in $\varphi$ belong to $CL(\varphi)$ and $\varphi$ can be transformed into PF form in this case.

\item[--] $\varphi$ is $\mu X.\phi$: let $p_X$ be an atomic proposition where $\llbracket p_X\rrbracket^{\mathcal{K}}= \llbracket \mu X.\phi \rrbracket^{\mathcal{K}}$ w.r.t. a certain linear time structure $\mathcal{K}$. As a result, $\phi[p_X/X]$ can be treated as a closed formula. By induction hypothesis, $\phi[p_X/X]$ can be transformed into PF form: \[\phi[p_X/X]\equiv\bigvee_{i=1}^{n}(\phi_{p_i}\wedge\bigcirc\phi_{f_i}[p_X/X])\]
Due to the restriction of guarded form, $p_X$ can only appear in the future part of the above PF form. Suppose
\[U_1=\{\phi_{f_1},\ldots,\phi_{f_m}\},\; U_2=\{\phi_{f_{m+1}},\ldots,\phi_{f_n}\}\]
where each $\phi_{f_j}\in U_1$ ($j\in\{1,\ldots,m\}$) does not contain $p_X$ while each $\phi_{f_k}\in U_2$ ($k\in\{m+1,\ldots,n\}$) contains $p_X$. By induction hypothesis, for each $\phi_{cj} \in Conj(\phi_{f_j})$ and $\phi_{ck}[p_X/X] \in Conj(\phi_{f_k})$, we have $\phi_{cj},\phi_{ck}[p_X/X] \in CL(\phi[p_X/X])$. Since $\mu X.\phi \equiv \phi[\mu X.\phi/X]$, then $\varphi$ can be converted into:
\[\varphi\equiv \phi[\mu X.\phi/X]\equiv \bigvee_{i=1}^{n}(\phi_{p_i}\wedge \bigcirc \phi_{f_i}[\mu X.\phi/p_X])\]
For each $\phi_{cj} \in CL(\phi[p_X/X])$, after the substitution of $\mu X.\phi$ for $p_X$, we can still have $\phi_{cj} \in CL(\phi[\mu X.\phi/p_X])$. Since $\phi[\mu X.\phi/p_X] \equiv \phi[\mu X.\phi/X]$ and $\phi[\mu X.\phi/X] \in CL(\varphi)$, then $\phi_{cj} \in CL(\varphi)$. For each $\phi_{ck}[p_X/X] \in CL(\phi[p_X/X])$, after the
substitution of $\mu X.\phi$ for $p_X$, we can further obtain
$\phi_{ck}[\mu X.\phi/p_X] \in CL(\phi[\mu X.\phi/p_X])$. Since
$\phi[\mu X.\phi/p_X] \equiv \phi[\mu X.\phi/X]$ and $\phi[\mu
X.\phi/X] \in CL(\varphi)$, we have $\phi_{ck}[\mu X.\phi/p_X] \in
CL(\varphi)$. Therefore, $\varphi$ can be transformed into PF form in this case.

\item[--] $\varphi$ is $\nu X.\phi$: this case can be proved similarly to the case $\varphi$ \emph{is} $\mu X.\phi$.
\end{itemize}

Thus, it can be concluded that every closed $\nu$TL formula can be transformed into PF form.
\hfill{$\Box$}
\subsection{Algorithm for Transforming a $\nu$TL Formula Into PF Form }
In this section we present algorithm \emph{PFTran} for transforming a closed $\nu$TL formula $\phi$ into PF form. The basic idea of the algorithm comes directly from the proof of Theorem \ref{PFFormThm}. Thus, its correctness can be ensured.
\begin{algorithm}[h]
\caption{PFTran($\phi$)} \label{PFTran}
\begin{algorithmic}[1]
\STATE \textbf{case}
\STATE ~~$\phi$ is $true$: \textbf{return} $true\wedge\bigcirc true$
\STATE ~~$\phi$ is $false$: \textbf{return} $false$
\STATE ~~$\phi$ is $\phi_{p}$ where $\phi_{p}\equiv\bigwedge_{h=1}^{n}\dot{p}_{h}$: \textbf{return} $\phi_{p} \wedge \bigcirc true$
\STATE ~~$\phi$ is $\phi_{p}\wedge\bigcirc\varphi$: \textbf{return} $\bigvee_{i} (\phi_{p} \wedge \bigcirc \varphi_i)$
\STATE ~~$\phi$ is $\bigcirc\varphi$: \textbf{return} $\bigvee_{i} (true \wedge \bigcirc \varphi_i)$
\STATE ~~$\phi$ is $\phi_1\vee\phi_2$: \textbf{return} $PFTran(\phi_1)\vee PFTran(\phi_2)$
\STATE ~~$\phi$ is $\phi_1\wedge\phi_2$: \textbf{return} $AND(PFTran(\phi_1), PFTran(\phi_2))$
\STATE ~~$\phi$ is $\sigma X.\varphi$: \textbf{return} $PFTran(\varphi[\sigma X.\varphi/X])$
\STATE \textbf{end case}
\end{algorithmic}
\end{algorithm}

In algorithm \emph{PFTran}, if $\phi$ is $true$ or $false$, the transformation is straightforward; if $\phi$ is $\phi_{p}$ where $\phi_{p}\equiv\bigwedge_{h=1}^{n}\dot{p}_{h}$, its PF form is $\phi_{p} \wedge \bigcirc true$; if $\phi$ is $\phi_{p}\wedge\bigcirc\varphi$, its PF form is $\bigvee_{i} (\phi_{p} \wedge \bigcirc \varphi_i)$; if $\phi$ is $\bigcirc\varphi$, its PF form is $\bigvee_{i} (true \wedge \bigcirc \varphi_i)$; if $\phi$ is $\phi_1\vee\phi_2$, the algorithm calls itself to transform $\phi_1$ and $\phi_2$ into PF form respectively; if $\phi$ is $\phi_1\wedge\phi_2$, the algorithm also calls itself first to transform $\phi_1$ and $\phi_2$ into PF form respectively and then converts $\phi_1\wedge\phi_2$ into PF form by algorithm \emph{AND}; if $\phi$ is $\sigma X.\varphi$, the algorithm transforms $\varphi[\sigma X.\varphi/X]$ into PF form.

\begin{algorithm}[h]
\caption{AND($\phi,\varphi$)}\label{ANDTran}
\begin{algorithmic}[1]
\IF{$\phi$ is of the form $\bigvee_{i}(\phi_i \wedge \bigcirc\phi_i')$ and $\varphi$ is of the form $\bigvee_{j}(\varphi_j \wedge \bigcirc\varphi_j')$}
\STATE \textbf{return} $\bigvee_{i}\bigvee_{j}(\phi_i\wedge\varphi_j\wedge\bigcirc(\phi_i'\wedge\varphi_j'))$
\ENDIF
\end{algorithmic}
\end{algorithm}

Algorithm \emph{AND} is used by \emph{PFTran} to deal with the $\wedge$ construct. Note that the inputs $\phi$ and $\varphi$ for \emph{AND} are both in PF form. Therefore, $\phi$ must be of the form $\bigvee_{i}(\phi_i \wedge \bigcirc\phi_i')$ while $\varphi$ of the form $\bigvee_{j}(\varphi_j \wedge \bigcirc\varphi_j')$.

In the following, we use an example to demonstrate how to transform a closed $\nu$TL formula into PF form by means of algorithm \emph{PFTran}.
\begin{Exm}
Transforming formula $\nu X.(r \wedge \bigcirc X) \wedge \mu Y.(q \vee p\wedge\bigcirc Y)$ into PF form by algorithm PFTran.
\end{Exm}
\[
\begin{array}{ll}
\,&PFTran(\nu X.(r \wedge \bigcirc X) \wedge \mu Y.(q \vee p\wedge\bigcirc Y))\\
\equiv&AND(PFTran(\nu X.(r \wedge \bigcirc X)), PFTran(\mu Y.(q \vee p\wedge\bigcirc Y)))\\
\equiv&AND(PFTran(r \wedge \bigcirc \nu X.(r \wedge \bigcirc X)), PFTran(q \vee p\wedge\bigcirc \mu Y.(q \\
\,&\vee p\wedge\bigcirc Y)))\\
\equiv&AND(r \wedge \bigcirc \nu X.(r \wedge \bigcirc X), PFTran(q) \vee PFTran(p\wedge\bigcirc\mu Y.(q \\
\,&\vee p\wedge\bigcirc Y)))\\
\equiv&AND(r \wedge \bigcirc \nu X.(r \wedge \bigcirc X), q\wedge \bigcirc true \vee p\wedge\bigcirc \mu Y.(q \vee p\wedge\bigcirc Y\\
\,&))\\
\equiv&r \wedge q \wedge \bigcirc \nu X.(r \wedge \bigcirc X) \vee r\wedge p \wedge \bigcirc (\nu X.(r \wedge \bigcirc X) \wedge \mu Y.(q\vee p\\
\,& \wedge\bigcirc Y))
\end{array}
\]

PF form enables us to convert a formula $\phi$ into two parts: the present and future ones. The present part is a conjunction of atomic propositions or their negations in $\phi$, while the future part is a next formula consisting of the conjunction of formulas in $CL(\phi)$. To make $\phi$ satisfied, the present part should be satisfied at the current state while the future part at the next one. Further, we can repeat the transformation process by converting each formula in the future part into PF form, which inspires us to construct a graph, namely Present Future form Graph (PFG), for describing models of $\phi$. This will be discussed in the next section.
\section{Present Future Form Graph}\label{sectionPFGConstruction}
\subsection{Definition of PFG}
For a closed $\nu$TL formula $\phi$, the PFG of $\phi$, denoted by $G_{\phi}$, is a tuple $(N_{\phi},E_{\phi},n_0)$ where $N_{\phi}$ is a set of nodes, $E_{\phi}$ a set of directed edges, and $n_0$ the root node. Each node in $N_{\phi}$ is specified by the conjunction of formulas in $CL(\phi)$ while each edge in $E_{\phi}$ is identified by a triple $(\phi_0,\phi_e,\phi_1)$, where $\phi_0$, $\phi_1\in N_{\phi}$ and $\phi_e$ is the label of the edge from $\phi_0$ to $\phi_1$. An edge may be associated with a mark which is a subset of variables occurring in $\phi$.

\begin{Def}
For a given closed $\nu$TL formula $\phi$, $N_{\phi}$ and $E_{\phi}$ can be inductively defined by:
\end{Def}
\begin{itemize}
\item[1)] $n_0=\phi \in N_{\phi};$
\item[2)] For all $\varphi \in N_{\phi}\setminus \{false\}$, if $\varphi\equiv
\bigvee_{i=1}^{k}(\varphi_{p_i}\wedge\bigcirc
\varphi_{f_i})$, then $\varphi_{f_i}\in N_{\phi}$, $(\varphi,\varphi_{p_i},\varphi_{f_i})\in E_{\phi}$ for each $i$ ($1\leq i \leq k$).
\end{itemize}
\begin{figure}[tbp]
  \centering
  \includegraphics[scale=.6]{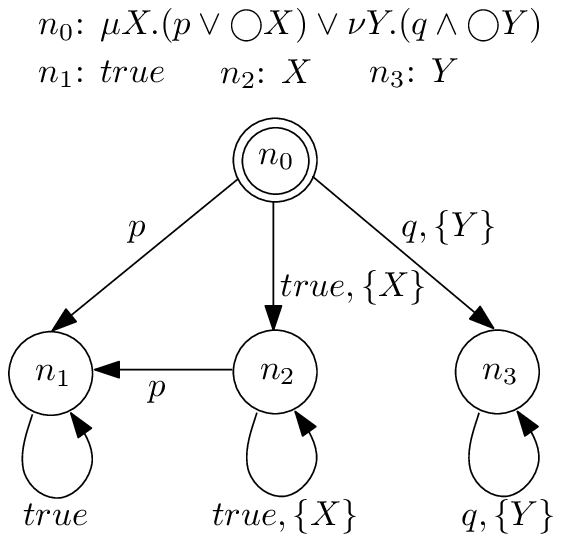}
  \caption{An example of PFG} \label{PFGExample}
\end{figure}

In a PFG, the root node is denoted by a double circle while each of other nodes by a single circle. Each edge is denoted by a directed arc with a label and also possibly a mark that connects two nodes. To simplify matters, we usually use variables to represent the corresponding fixpoint formulas occurring in a node. An example of PFG for formula $\mu X.(p\vee\bigcirc X) \vee \nu Y.(q\wedge\bigcirc Y)$ is depicted in Fig. \ref{PFGExample}. There are four nodes in the PFG where $n_0$ is the root node. $(n_0,q,n_3)$ is an edge with label being $q$ and mark being $\{Y\}$ while $(n_0,p,n_1)$ is an edge with label being $p$ and no mark.

\subsection{Marks in PFG}
From Fig. \ref{PFGExample} we can see that there may exist a path in a PFG, e.g. $n_0,true,(n_2,$ $true)^{\omega}$, which arises from the infinite unfolding of a least fixpoint formula. Thus, marks are useful in a PFG to keep track of the infinite unfolding problem for least fixpoint formulas when constructing the PFG.
\begin{Def}
Given a PFG $G_{\phi}$ and a node $\phi_m \in N_{\phi}$ where $\phi_m \equiv
\bigvee_{i=1}^{k}(\phi_{p_i}\wedge\bigcirc \phi_{f_i})$. The \emph{mark} of edge $(\phi_m,\phi_{p_i},\phi_{f_i})$ $(1 \leq i\leq k)$ is a set of variables $M_v$ such that for each $X \in M_v$, the fixpoint formula $\sigma X.\phi_X$ identified by $X$ appears as a subformula of $\phi_{f_i}$ and has not been unfolded by formula $\nu Y.\phi_Y$ where $Y$ is higher than $X$ in the PF form transformation process.
\end{Def}
We use the notion of $\mu$-signatures to demonstrate how to add marks to a PFG. Intuitively, a variable $X$ is added to a mark in a PF form transformation process if the unfolding of the corresponding formulas does not increase the $\mu$-signature w.r.t. $X$. As a result, we can use marks to detect the infinite descending chains of $\mu$-signatures.
\begin{figure}[bp]
  \centering
  \includegraphics[scale=.6]{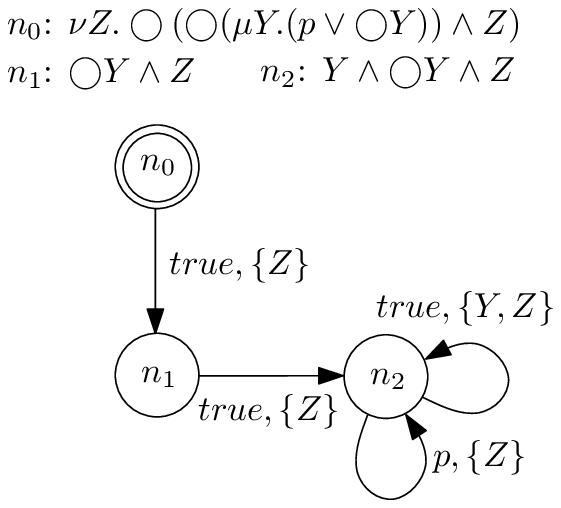}
  \caption{PFG of $\nu Z.\bigcirc(\bigcirc(\mu Y.(p\vee\bigcirc Y))\wedge Z)$} \label{UnfoldByItself}
\end{figure}

When transforming a formula into its PF form, the occurrence of a fixpoint formula $\sigma X.\phi_X$ in the future part $\phi_{f_i}$ may be caused by the unfolding of: (I) itself, (II) a least fixpoint formula $\mu Y.\phi_Y$ where $Y$ is higher than $X$, or (III) a greatest fixpoint formula $\nu Z.\phi_Z$ where $Z$ is higher than $X$. According to Lemma 3.5 in \cite{streett1989automata}, the $\mu$-signature w.r.t. $X$ does not increase unless the case III happens. For example, as shown in Fig. \ref{UnfoldByItself}, when node $n_0$ is transformed into PF form: $n_0\equiv true\wedge \bigcirc n_1$, the occurrence of $\mu Y.(p\vee\bigcirc Y)$ in $n_1$ is due to the unfolding of $\nu Z.\bigcirc(\bigcirc(\mu Y.(p\vee\bigcirc Y))\wedge Z)$, hence $Y$ does not exist in the mark of edge $(n_0,true,n_1)$.

Note that cases I and II, or I and III (e.g. the occurrence of $Y$ in the mark of edge $(n_2,true,n_2)$ in Fig. \ref{UnfoldByItself}) can occur simultaneously. If that happens, we can see that the $\mu$-signature w.r.t. $X$ still does not increase. In particular, cases II and III cannot happen simultaneously since $X$ is bound exactly once.

Given a formula $\sigma X.\phi$, to construct its PFG sometimes we need to deal with a formula of the form $\bigwedge_{i=1}^{n}\sigma_i X_i.\phi_i$, where each $\sigma_i X_i.\phi_i \in CL(\sigma X.\phi)$ and $i<j$ implies $X_i$ is higher than $X_j$, in a PF form transformation process. It is straightforward that the unfolding of $\sigma_n X_n.\phi_n$ ensures that the $\mu$-signature w.r.t. each $X_i$ does not increase after the transformation despite the value of each $\sigma_i$.

In the following, we use a simple example to illustrate how the marks work.

\begin{Exm}\label{MarkOnPFGExample}
Tracing the infinite unfolding of $\mu X.(p\vee\bigcirc X)$ using marks.
\end{Exm}
\begin{figure}[tbp]
  \centering
  \includegraphics[scale=.6]{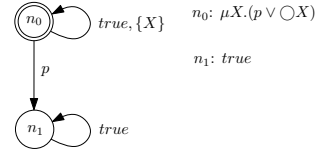}
  \caption{How the marks work} \label{MarkExm}
\end{figure}
The PF form of $\mu X.(p\vee\bigcirc X)$ is: $\mu X.(p\vee\bigcirc X)\equiv p\wedge\bigcirc true \vee true \wedge \bigcirc\mu X.(p\vee\bigcirc X)$. The second disjunct of the PF form leads to the generation of edge $(n_0,true,n_0)$ in Fig. \ref{MarkExm}. Then mark $\{X\}$ is added accordingly since $\mu X.(p\vee\bigcirc X)$ appears in the future part of the PF form and has been unfolded by itself in the PF form transformation process. Moreover, it is easy to see that all other edges have no marks. Formula $\mu X.(p\vee\bigcirc X)$ indicates that the atomic proposition $p$ finally holds somewhere and therefore path $(n_0,true)^{\omega}$ does not characterize a model. Actually, $(n_0,true)^{\omega}$ is generated by the infinite unfolding of $\mu X.(p\vee\bigcirc X)$ and the infinite occurrence of mark $\{X\}$ on this path describes exactly an infinite descending chain of $\mu$-signatures for $\mu X.(p\vee\bigcirc X)$. This is why we need to use marks.

\subsection{Paths in PFG}
A \emph{path} $\Pi$ in a PFG $G_{\phi}$ is an infinite alternate sequence of nodes and edges departing from the root node. In the following, we show how to establish the relationship between paths in PFG and linear time structures.

\begin{figure}[tbp]
  \centering
  \includegraphics[scale=.6]{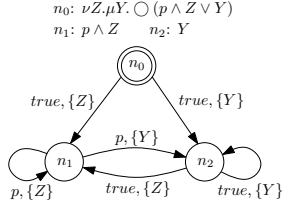}
  \caption{Paths in PFG} \label{PathExm}
\end{figure}

Let $Atom(\bigwedge_{i=1}^{m}\dot{q_i})$ denote the set of atomic propositions or their negations appearing in formula $\bigwedge_{i=1}^{m}\dot{q_i}$. Given a path $\Pi=\phi_0,\phi_{e0},\phi_1,\phi_{e1},\ldots$ in a PFG, we can obtain a corresponding linear time structure $Atom(\phi_{e0}),Atom(\phi_{e1}),\ldots$.
\begin{Exm}\label{PathsInPFGExm}
Paths in Fig. \ref{PathExm}.
\end{Exm}
\begin{itemize}
\item[1)] Path $n_0,true,(n_1,p)^\omega$ corresponds to the linear time structure $\{true\}\{p\}^{\omega}$.
\item[2)] Path $n_0,true,(n_2,true,n_1,p)^\omega$ corresponds to the linear time structure $\{true\}(\{true\}\{p\})^\omega$.
\end{itemize}

Actually, each node in the PFG $G_{\phi}$ of formula $\phi$ corresponds precisely to a consistent subset of $CL(\phi)$. In other words, each path in $G_{\phi}$ characterizes a pre-model \cite{vardi1988temporal,streett1989automata} of $\phi$. A pre-model is almost a model except that it ignores the infinite unfolding problem for least fixpoint formulas. We can distinguish real models from all pre-models using marks.
\subsection{Algorithm for Constructing PFG}
Given a closed $\nu$TL formula $\phi$, the whole process of constructing its PFG $G_\phi$ is presented in Algorithm \ref{PFGConstruction}.
\begin{algorithm}[h]
\caption{PFGCon($\phi$)}\label{PFGConstruction}
\begin{algorithmic}[1]
\STATE $n_0=\phi$, $N_{\phi}=\{n_0\}$, $E_{\phi}=\emptyset$, isHandled[$n_0$] = 0
\WHILE{there exists $\varphi \in N_{\phi}\setminus\{false\}$ and isHandled[$\varphi$] = 0}
\STATE isHandled[$\varphi$] = 1
\STATE $\varphi$ = PFTran($\varphi$)  \quad/*suppose $\varphi=\bigvee_{i=1}^{k}(\varphi_{p_i}\wedge\bigcirc
\varphi_{f_i})$*/
\FOR{i = 1 to k}
\IF{$\varphi_{p_i}$ is not $false$}
\STATE $E_{\phi}=E_{\phi} \cup \{(\varphi,\varphi_{p_i},\varphi_{f_i})\}$ \quad/*adding edges*/
\STATE AddMark($(\varphi,\varphi_{p_i},\varphi_{f_i})$) /*obtaining the corresponding marks*/
\IF{$\varphi_{f_i}\notin N_{\phi}$}
\STATE $N_{\phi}=N_{\phi}\cup \{\varphi_{f_i}\}$ \quad/*adding nodes*/
\IF{$\varphi_{f_i}$ is not $false$}
\STATE isHandled[$\varphi_{f_i}$] = 0 \quad/*$\varphi_{f_i}$ is a new node which needs to be handled*/
\ELSE
\STATE isHandled[$\varphi_{f_i}$] = 1 \quad/*$\varphi_{f_i}$ does not need to be handled*/
\ENDIF
\ENDIF
\ENDIF
\ENDFOR
\ENDWHILE
\FORALL{$\varphi\in N_{\phi}$ with no outgoing edge}
\STATE $N_{\phi}=N_{\phi}\setminus \{\varphi\}$ \quad/*eliminating redundant nodes and the relative edges*/
\STATE $E_{\phi}=E_{\phi}\setminus \bigcup_{i}\{(\varphi_i,\varphi_e,\varphi)\}$
\ENDFOR
\RETURN $G_{\phi}$
\end{algorithmic}
\end{algorithm}

The algorithm takes $\phi$ as input and returns $G_{\phi}$. First, $n_0$ is assigned to $\phi$. $N_{\phi}$ and $E_{\phi}$ are initialized to \{$n_0$\} and empty, respectively. Further, the algorithm repeatedly converts an unhandled formula $\varphi \in N_{\phi}$ into PF form by algorithm \emph{PFTran} and then adds the corresponding nodes and edges to $N_{\phi}$ and $E_{\phi}$, respectively, until all formulas in $N_{\phi}$ have been handled. $isHandled[]$ is used to indicate whether a formula has been handled. If $isHandled[\varphi]=0$, $\varphi$ needs further to be handled; otherwise, $\varphi$ has been handled or there is no need to handle it. Function $AddMark$ is utilized to mark an edge with a subset of variables occurring in $\phi$ by distinguishing appropriate fixpoint formulas from all fixpoint formulas contained in the future part of a certain PF form.
\begin{algorithm}[h]
\caption{AddMark($(\varphi,\varphi_{p_i},\varphi_{f_i})$)}\label{AddMark}
\begin{algorithmic}[1]
\FOR{each conjunct $\varphi_c$ of $\varphi_{f_i}$}
\IF{$\varphi_c$ is of the form $\bigcirc^{n}\sigma X.\varphi_X$ and flag[$\sigma X.\varphi_X$] = 0}
\STATE $M_i = M_i \cup \{X\}$ \quad/*$M_i$ represents the mark of edge $(\varphi,\varphi_{p_i},\varphi_{f_i})$*/
\ENDIF
\ENDFOR
\end{algorithmic}
\end{algorithm}

The input for algorithm $AddMark$ is an edge $(\varphi,\varphi_{p_i},\varphi_{f_i})$ in $G_{\phi}$. In the algorithm, $flag[]$ is employed to denote whether a fixpoint formula $\varphi_{fix}$ appearing in the future part of a PF form has been unfolded by a greatest fixpoint formula $\nu Y.\varphi_Y$ where $Y$ is higher than the variable identifying $\varphi_{fix}$ in the PF form transformation process. If $flag[\varphi_{fix}]=1$, $\varphi_{fix}$ is unfolded by $\nu Y.\varphi_Y$; otherwise, it is unfolded by itself or a least fixpoint formula. For any fixpoint subformula $\sigma Z.\varphi_{sub}$ of $\varphi$, $flag[\sigma Z.\varphi_{sub}]$ is assigned to 0 before $\varphi$ is transformed into PF form. For the input $(\varphi,\varphi_{p_i},\varphi_{f_i})$, $AddMark$ checks each conjunct $\varphi_c$ of $\varphi_{f_i}$. If $\varphi_c$ is in the form $\bigcirc^{n}\sigma X.\varphi_X$ ($n\geq 0$) and $flag[\sigma X.\varphi_X]=0$, $X$ is added to $M_i$. Here $\bigcirc^{n}$ represents the consecutive occurrence of $\bigcirc$ operators for $n$ times and $M_i$ represents the mark of the edge $(\varphi,\varphi_{p_i},\varphi_{f_i})$.

Additionally, it is worth pointing out that, throughout the construction of $G_{\phi}$, a \emph{false} node (e.g. $p\wedge\neg p$) may be generated which corresponds to an inconsistent subset of $CL(\phi)$. Such kind of nodes have no successor and are redundant. We use the \textit{for loop} in Line 20 of \emph{PFGCon} to remove those redundant nodes as well as the relative edges.

\begin{Exm}
Constructing the PFG of formula $\mu X.(p\vee\bigcirc X)\vee \nu Y.(q\wedge\bigcirc Y)$ by algorithm PFGCon.
\end{Exm}
As depicted in Fig. \ref{PFGExample}, at the very beginning, the
root node $\mu X.(p\vee\bigcirc X)\vee \nu Y.(q\wedge\bigcirc Y)$ is
created and denoted by $n_0$; then we transform $\mu
X.(p\vee\bigcirc X)\vee \nu Y.(q\wedge\bigcirc Y)$ into PF form:
\[\mu X.(p\vee\bigcirc X)\vee \nu Y.(q\wedge\bigcirc Y)\equiv p\wedge\bigcirc true \vee true \wedge \bigcirc \mu X.(p\vee\bigcirc X)\]
\[~~~~~~~~~~~~~~~~~~~~~~~~\vee q\wedge\bigcirc \nu Y.(q\wedge\bigcirc Y)\]
Accordingly, nodes $true$, $\mu X.(p\vee\bigcirc X)$ and $\nu Y.(q\wedge\bigcirc Y)$ are created and denoted respectively by $n_1$, $n_2$ and $n_3$. Meanwhile, edges $(n_0,p,n_1)$, $(n_0,true,n_2)$ and $(n_0,q,n_3)$ are created among which $(n_0,true,n_2)$ is marked with $\{X\}$ and $(n_0,q,n_3)$ with $\{Y\}$. Further, $true$ is transformed into PF form: $true\equiv true\wedge\bigcirc true$.
Thus, edge $(n_1,true,n_1)$ is created. After that, $\mu X.(p\vee\bigcirc
X)$ is transformed into PF form: $\mu X.(p\vee\bigcirc X)\equiv p\wedge\bigcirc true \vee true \wedge \bigcirc \mu X.(p\vee\bigcirc X)$. Hence, edges $(n_2,p,n_1)$ and $(n_2,true,n_2)$ are created where
$(n_2,$ $true,n_2)$ is marked with $\{X\}$. Finally, $\nu
Y.(q\wedge\bigcirc Y)$ is transformed into PF form: $\nu Y.(q\wedge\bigcirc Y)\equiv q\wedge\bigcirc \nu Y.(q\wedge\bigcirc Y)$. Accordingly, edge $(n_3,q,n_3)$ is created with the mark being $\{Y\}$, and the whole construction process terminates.

\subsection{Finiteness of PFG}
In the PFG $G_{\phi}$ of formula $\phi$ generated by algorithm \emph{PFGCon}, $N_{\phi}$ and $E_{\phi}$ are produced by repeatedly transforming the unhandled nodes into PF form. Since each node in $N_{\phi}$ is the conjunction of formulas in $CL(\phi)$, the following corollary is easily obtained.
\begin{Cor}\label{PFGFinitenessCor}
For any closed $\nu$TL formula $\phi$, the number of nodes in $G_{\phi}$ is bounded by $2^{O(|\phi|)}$, and the number of edges in $G_{\phi}$ is bounded by $2^{O(|\phi|)}\cdot 2^{O(n_p)} \cdot 2^{O(n_v)} \cdot 2^{O(|\phi|)}$ (which is also $2^{O(|\phi|)}$), where $n_p$ and $n_v$ denote the number of atomic propositions and fixpoint subformulas occurring in $\phi$ respectively.
\end{Cor}
\section{Decision Procedure Based on PFG}\label{sectionDecisionProcedure}
In this section we show how to find a model for a given closed $\nu$TL
formula $\phi$ from its PFG $G_{\phi}$. In fact, each outgoing edge of a node in $G_{\phi}$ amounts to a possible choice prescribed by an underlying choice function \cite{streett1989automata}. Since a node cannot have multiple choices simultaneously, we restrict ourselves here only to paths ending with simple loops in $G_{\phi}$. Let $\Pi$ be a path in $G_{\phi}$, for convenience, we use $LES(\Pi)$ to denote the set of edges appearing in the loop part of $\Pi$, $Mark(E)$ the mark of edge $E$, $LMS(\Pi)$ the set of all $\mu$-variables occurring in each $Mark(E_l)$ where $E_l \in LES(\Pi)$.

\subsection{$\nu$-path}
\begin{figure}[bp]
  \centering
  \includegraphics[scale=.55]{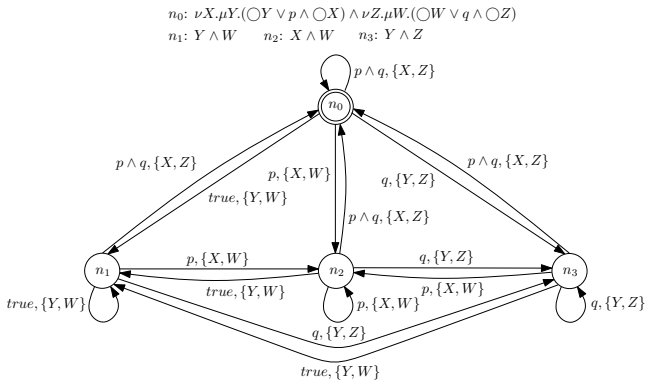}
  \caption{$\nu$-paths in PFG} \label{NuPathsExm}
\end{figure}

Here we present the notion of $\nu$-\emph{paths} which will play a vital role in obtaining the PFG-based decision procedure for $\nu$TL.
\begin{Def}\label{NuPathDef}
Given a PFG $G_{\phi}$ and a path $\Pi$ in $G_{\phi}$, we call $\Pi$ a $\nu$-\emph{path} iff for each $X \in LMS(\Pi)$, an edge $E \in LES(\Pi)$ can be found such that $X \notin Mark(E)$ and there exists no $X'\in Mark(E)$ where $X \lhd X'$.
\end{Def}
\begin{Exm}\label{NuPathsInPFGExm}
$\nu$-paths in Fig. \ref{NuPathsExm}.
\end{Exm}
\begin{itemize}
\item[1)] $\Pi_1$: $(n_0,p\wedge q)^\omega$. $\Pi_1$ is a $\nu$-path since $LMS(\Pi_1)=\emptyset$.
\item[2)] $\Pi_2$: $n_0,true,(n_1,true)^\omega$. We have $LES(\Pi_2)=\{(n_1,true,$ $n_1)\}$ and $LMS(\Pi_2)=\{Y,W\}$. For the first variable $Y\in LMS(\Pi_2)$, we cannot find an edge from $LES(\Pi_2)$ whose mark does not contain $Y$. So $\Pi_2$ is not a $\nu$-path.
\item[3)] $\Pi_3$: $n_0,p,(n_2,true,n_1,p)^\omega$. We have $LES(\Pi_3)=\{(n_2,true,$ $n_1),(n_1,p,n_2)\}$ and $LMS(\Pi_3)=\{Y,W\}$. For the first variable $Y \in LMS(\Pi_3)$, we can find an edge $(n_1,p,n_2) \in LES(\Pi_3)$ whose mark does not contain $Y$ and any variable depending on $Y$. However, for the second varible $W \in LMS(\Pi_3)$, we cannot find an edge from $LES(\Pi_3)$ whose mark does not contain $W$. Therefore, $\Pi_3$ is not a $\nu$-path.
\item[4)] $\Pi_4$: $(n_0,q,n_3,p\wedge q)^\omega$. We have $LES(\Pi_4)=\{(n_0,q,n_3),(n_3,$ $p\wedge q,n_0)\}$ and $LMS(\Pi_4)$ $=\{Y\}$. For the only variable $Y \in LMS(\Pi_4)$, we can find an edge $(n_3,p\wedge q,n_0) \in LES(\Pi_4)$ whose mark does not contain $Y$ and any variable depending on $Y$. Thus, $\Pi_4$ is a $\nu$-path.
\item[5)] $\Pi_5$: $n_0,p,(n_2,true,n_1,q,n_3,true,n_1,p)^\omega$. We have $LES(\Pi_5)=\{(n_2,true,n_1),(n_1,q,n_3),(n_3,true,n_1),(n_1,p,$ $n_2)\}$ and $LMS(\Pi_5)=\{Y,W\}$. For the first variable $Y \in LMS(\Pi_5)$, we can find an edge $(n_1,p,n_2) \in LES(\Pi_5)$ whose mark does not contain $Y$ and any variable depending on $Y$. Further, for the second varible $W \in LMS(\Pi_5)$, we can find an edge $(n_1,q,n_3) \in LES(\Pi_5)$ whose mark does not contain $W$ and any variable depending on $W$. Therefore, $\Pi_5$ is a $\nu$-path.
\end{itemize}

Regarding the notion of $\nu$-paths, the following theorem is formalized.
\begin{Thm}\label{SatisfyThm}
A closed $\nu$TL formula $\phi$ is satisfiable iff a $\nu$-path can be found in $G_{\phi}$.
\end{Thm}
\textit{Proof.} ($\Rightarrow$) Suppose $\phi$ is satisfiable and no $\nu$-path exists in $G_{\phi}$. In this case, for any path $\Pi_1$ in $G_{\phi}$, there exists at least one $X \in LMS(\Pi_1)$ such that for each edge $E_1 \in LES(\Pi_1)$, either $X \in Mark(E_1)$ or $X' \in Mark(E_1)$, where $X \lhd X'$.

As a result, we can obtain the following sequence of variables according to the sequence of marks in the loop part of $\Pi_1$:
\[X,X_1,X_2,\ldots,X_n,X\]
where each $X_i$ ($1\leq i \leq n$) is either $X$ itself or a variable depending on $X$.

Further, according to the sequence of variables above, we can acquire the following sequence of fixpoint formulas:
\[\mu X.\phi_{X},\sigma X_1.\phi_{1},\sigma X_2.\phi_{2},\ldots,\sigma X_n.\phi_{n},\mu X.\phi_{X}\]
Here each $\sigma X_i.\phi_{i}$ is identified by $X_i$ and $\mu X.\phi_{X}$ by $X$. Since each $X_i$ is either $X$ or a variable depending on $X$, $\mu X.\phi_{X}$ must appear as a subformula of each $\sigma X_i.\phi_{i}$. According to the way the marks are added, it can be seen that the above sequence describes exactly an infinite descending chain of $\mu$-signatures w.r.t. $X$. By the well-foundedness of $\mu$-signatures we can derive that $\Pi_1$ does not characterize a model of $\phi$. This contradicts the premise that $\phi$ is satisfiable. Therefore, if $\phi$ is satisfiable, there exists at least one $\nu$-path in $G_{\phi}$.

($\Leftarrow$) Let $\Pi_2$ be a $\nu$-path in $G_{\phi}$.

When $LMS(\Pi_2)$ is empty, no infinite descending chain of $\mu$-signatures on $\Pi_2$ can be detected. Consequently, $\Pi_2$ characterizes a model of $\phi$.

When $LMS(\Pi_2)$ is not empty, we have that for each $Y \in LMS(\Pi_2)$, an edge $E_2 \in LES(\Pi_2)$ can be found such that $Y \notin Mark(E_2)$ and there exists no $Y'\in Mark(E_2)$ where $Y \lhd Y'$. Subsequently, for each sequence of variables relevant to $Y$ obtained according to the sequence of marks in the loop part of $\Pi_2$:
\[Y,Y_1,Y_2,\ldots,Y_m,Y\]
We can obtain the following sequence of fixpoint formulas:
\[\mu Y.\phi_{Y},\sigma Y_1.\phi_{1},\sigma Y_2.\phi_{2},\ldots,\sigma Y_m.\phi_{m},\mu Y.\phi_{Y}\]
where there must exist a formula $\sigma Y_j.\phi_{j}$ ($1\leq j \leq m$) in which $\mu Y.\phi_{Y}$ does not appear as a subformula. Similarly, we have that $\Pi_2$ characterizes a model of $\phi$ according to the well-foundedness of $\mu$-signatures w.r.t. $Y$. It follows that when there exists a $\nu$-path in $G_{\phi}$, $\phi$ is satisfiable.
\hfill{$\Box$}

Consequently, we reduce the satisfiability problem of $\nu$TL formulas to a $\nu$-path searching problem from a PFG.
\begin{Exm}\label{ModelsInPFGExm}
Checking satisfiability of the following formulas.
\begin{itemize}
\item[(1)] $\nu Z.(\mu X.(\bigcirc X\vee \nu Y.(p\wedge\bigcirc Y))\wedge\bigcirc Z)$
\item[(2)] $\nu X.(p\wedge \bigcirc X) \wedge \nu Y.(\neg p\wedge \bigcirc Y)$
\item[(3)] $\mu X.(\mu Y.(p\wedge\bigcirc Y)\vee\bigcirc X)$
\end{itemize}
\end{Exm}
\begin{figure}[tbp]
  \centering
  \includegraphics[scale=.6]{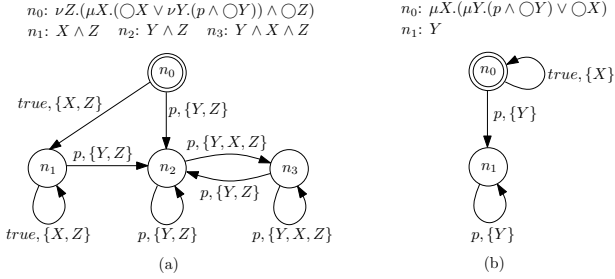}
  \caption{Examples of PFGs for satisfiability checking} \label{SatCheckingExm}
\end{figure}
For formula (1), as depicted in Fig. \ref{SatCheckingExm} (a), since a $\nu$-path $n_0,p,(n_2,p,n_3,p)^{\omega}$ can be found in its PFG, it is satisfiable. For formula (2), as its PFG is empty and contains no $\nu$-path, it is unsatisfiable. For formula (3), as depicted in Fig. \ref{SatCheckingExm} (b), no $\nu$-path exists in its PFG and hence it is unsatisfiable.
\subsection{Implementation of the Decision Procedure}
Based on Theorem \ref{SatisfyThm}, a PFG-based decision procedure, algorithm \textit{PFGSAT}, for checking satisfiability of $\nu$TL formulas is derived.
\begin{algorithm}[h]
\caption{PFGSAT($\phi$)}
\begin{algorithmic}[1]
\STATE $G_{\phi}$ = PFGCon$(\phi)$
\IF{$G_{\phi}$ is empty}
\RETURN unsatisfiable
\ENDIF
\STATE Tarjan($G_{\phi}$, $n_0$)
\FOR {each scc $\in$ sccs}
\STATE SCCNuSearch($v$, scc) \quad/*$v$ is an arbitrary node in scc*/
\ENDFOR
\RETURN unsatisfiable
\end{algorithmic}
\end{algorithm}

The algorithm takes a closed $\nu$TL formula $\phi$ as input and returns the result whether $\phi$ is satisfiable in the end. To do so, the PFG, $G_{\phi}$, of $\phi$ is constructed first. Next, it checks whether $G_{\phi}$ is empty: if so, $\phi$ is unsatisfiable since no $\nu$-path can be found in $G_{\phi}$; otherwise, the algorithm will try to find a $\nu$-path in $G_{\phi}$. Further, algorithm \textit{Tarjan} is employed to compute all strongly connected components (SCCs) in $G_{\phi}$. Finally, the algorithm checks whether there exists a loop in some SCC which corresponds to a $\nu$-path by algorithm \textit{SCCNuSearch}: if so, \textit{SCCNuSearch} will return that $\phi$ is satisfiable; otherwise, $\phi$ is unsatisfiable.

\begin{algorithm}[h]
\caption{Tarjan($G_{\phi}$, $v$)}\label{AlgorithmTarjan}
\begin{algorithmic}[1]
\STATE dfn[v] = low[v] = ++index
\STATE visit[v] = 1
\STATE Stack.push(v)
\FOR {each edge e $\in E_{\phi}$ where src[e] = v}
\IF{visit[tgt[e]] = 0}
\STATE Tarjan($G_{\phi}$, tgt[e])
\STATE low[v] = min\{low[v], low[tgt[e]]\}
\ELSE
\IF{tgt[e] is in Stack}
\STATE low[v] = min\{low[v], dfn[tgt[e]]\}
\ENDIF
\ENDIF
\ENDFOR
\IF{dfn[v] = low[v]}
\STATE subGraph scc
\REPEAT
\STATE u = Stack.top()
\STATE Stack.pop()
\STATE scc.push\_back(u)
\UNTIL{v = u}
\STATE sccs.push\_back(scc)
\ENDIF
\end{algorithmic}
\end{algorithm}

\textbf{SCC Computation.} Tarjan algorithm \cite{tarjan73} presented in Algorithm \ref{AlgorithmTarjan} is a classical algorithm for computing SCCs in a graph based on depth-first search (DFS). The algorithm takes a PFG $G_{\phi}$ and a node $v$ in $G_{\phi}$ as inputs and acquires all SCCs in $G_{\phi}$. \textit{dfn}[u] is employed to represent the timestamp of a given node $u$ indicating the number of nodes which have been visited before $u$ is visited, while \textit{low}[u] the timestamp of the earliest node reachable from $u$ or subtrees of $u$. Also, we use \emph{visit}[] to denote whether a node $u$ has been visited. If \emph{visit}[u] = 1, $u$ has already been visited; otherwise, $u$ has not been visited yet. For each node $u$ in $G_{\phi}$, \emph{visit}[u] is initialized to $0$. \emph{src}[] and \emph{tgt}[] are utilized to obtain the source and target nodes of an edge, respectively.
\begin{algorithm}[h]
\caption{SCCNuSearch($v$, scc)}
\begin{algorithmic}[1]
\STATE NS.push\_back(v)
\FOR {each edge e in scc}
\IF{src[e] = v and visit[e] = 0}
\STATE ES.push\_back(e)
\STATE visit[e] = 1
\IF{isLoop(tgt[e], pos)}
\STATE TES.assign(ES.begin() + pos, ES.end())
\IF{isNuPath(TES)}
\RETURN satisfiable
\ENDIF
\STATE ES.pop\_back()
\ELSE
\STATE SCCNuSearch(tgt[e], scc)
\ENDIF
\ENDIF
\ENDFOR
\IF{ES.size() $>$ 0}
\STATE ES.pop\_back()
\ENDIF
\STATE NS.pop\_back()
\end{algorithmic}
\end{algorithm}

\textbf{Path Construction.} Given an SCC \textit{scc} in a PFG $G_{\phi}$ and an arbitrary node $v$ in \textit{scc}, we use algorithm \emph{SCCNuSearch} to build a path which is likely to correspond to a $\nu$-path in $G_{\phi}$. Two global variables, \emph{ES} and \emph{NS}, are used in the algorithm. \emph{ES} is a vector which stores the sequence of edges aiming to construct a path ending with a loop. \emph{NS} is also a vector storing the sequence of nodes corresponding to \emph{ES}. In addition, \emph{src}[] and \emph{tgt}[] are employed to obtain the source and target nodes of an edge, respectively. The algorithm uses \emph{visit}[] to indicate whether an edge $e$ has been visited. If \emph{visit}[e] = 1, $e$ has already been visited; otherwise, $e$ has not been visited yet. For each edge $e$ in $G_{\phi}$, \emph{visit}[e] is initialized to $0$. \emph{isLoop} and \emph{isNuPath} are two boolean functions. \emph{isLoop} determines whether a node $u$ exists in \emph{NS} and obtains, if so, the position of $u$ in \emph{NS}. \emph{isNuPath} determines whether a sequence of edges corresponds to a $\nu$-path.

\begin{algorithm}[h]
\caption{isLoop($v$, pos)}
\begin{algorithmic}[1]
\STATE counter = 0
\FOR {each node u $\in$ NS}
\STATE counter++
\IF{u = v}
\STATE pos = counter
\RETURN true
\ENDIF
\ENDFOR
\RETURN false
\end{algorithmic}
\end{algorithm}

In algorithm \emph{SCCNuSearch}, $v$ is added to \emph{NS} first. After that, for each unvisited edge $e$ in \textit{scc} whose source node is $v$, the algorithm adds it to \emph{ES} and assigns \emph{visit}[e] to 1. Then, it determines whether \emph{tgt}[e] exists in \emph{NS} by means of algorithm \emph{isLoop}. If the output of \emph{isLoop} is $true$, there exists a loop in \emph{ES} and we use \emph{TES} to store the loop of \emph{ES}. Further, algorithm \emph{isNuPath} is called to decide whether \emph{TES} corresponds to a $\nu$-path. If the output of \emph{isNuPath} is $true$, the given formula is satisfiable and the algorithm terminates; otherwise, the last edge in \emph{ES} is removed and a new \textit{for loop} begins in order to search for another unvisited edge from \textit{scc} whose source node is $v$ to establish a new path. In case the output of \emph{isLoop} is $false$, which means the current \emph{ES} cannot construct a path ending with a loop, the algorithm calls itself and tries to build new paths from node \emph{tgt}[e]. If the conditional statement in Line 3 is never satisfied, i.e., any edge in \textit{scc} with $v$ being its source node has been visited, $v$ is removed from \emph{NS}. Note that if the size of \emph{ES} is greater than $0$ when the loop terminates, we need to remove the last edge in \emph{ES} generated by the next level of recursion.
\begin{algorithm}[h]
\caption{isNuPath(TES)}
\begin{algorithmic}[1]
\FOR {each edge e $\in$ TES}
\IF{$X \in$ Mark(e) and $X$ is a $\mu$-variable}
\STATE MS = MS $\cup$ $\{X\}$
\ENDIF
\ENDFOR
\FOR{each $V$ $\in$ MS}
\FOR{each $e'$ $\in$ TES}
\IF{$V\in$ Mark($e'$) or $V'\in$ Mark($e'$) where $V \lhd V'$}
\STATE c++
\STATE \textbf{continue}
\ELSE
\STATE c = 0
\STATE \textbf{break}
\ENDIF
\ENDFOR
\IF{c $>$ 0}
\RETURN false
\ENDIF
\ENDFOR
\RETURN true
\end{algorithmic}
\end{algorithm}

\textbf{$\nu$-path Determination.} Given a sequence of edges \emph{TES}, we use algorithm \textit{isNuPath} to determine whether \emph{TES} corresponds to a $\nu$-path. The algorithm uses \emph{MS} to denote the set of all $\mu$-variables appearing in each \emph{Mark(e)} where $e\in$ \emph{TES}. $c$ is a counter calculating how many edges in \emph{TES} have been handled by the \textit{for loop} in Line 7 of algorithm $isNuPath$ and initialized to $0$.

For the input \emph{TES}, the algorithm first computes the set of $\mu$-variables \emph{MS}. For each edge $e\in$ \emph{TES}, if there exists a $\mu$-variable $X\in Mark(e)$, $X$ is added to \emph{MS}. In this way, \emph{MS} can be obtained. Subsequently, to confirm whether \emph{TES} corresponds to a $\nu$-path, we need to seek out an edge $e'\in$ \emph{TES} for each $V\in MS$ such that $V \notin$ \emph{Mark(e$'$)}, and meanwhile there exists no $V' \in$ \emph{Mark(e$'$)} such that $V \lhd V'$. Further, for the conditional statement in Line 8, if the else-branch can never be performed, $c$ will be equal to the size of \emph{TES} when the inner \textit{for loop} terminates. Consequently, the condition in Line 16 is satisfied. That is, $false$ is returned by the algorithm, which indicates that \emph{TES} does not correspond to a $\nu$-path. When the else-branch is executed, $c$ will be assigned to 0 and then we use the break statement to jump out of the inner \textit{for loop}. In this case, the condition in Line 16 cannot be satisfied and the outer \textit{for loop} proceeds to deal with the next $\mu$-variable in \emph{MS}.

Note that if the else-branch can always be executed for each $V\in MS$, the algorithm will finally return $true$, which means \emph{TES} indeed corresponds to a $\nu$-path.

All the above-mentioned algorithms have been implemented in C++. In what follows we exhibit several PFGs generated by our tool.

\begin{Exm}\label{PFGByToolExm}
PFGs generated by our tool.
\begin{itemize}
\item[I.] $\mu X.\nu Y.(\bigcirc X \vee p \wedge \bigcirc Y) \wedge \nu Z.\mu W.(\bigcirc W \vee q \wedge \bigcirc Z)$
\item[II.] $\mu X.\mu Y.(q \wedge \bigcirc X \vee p \wedge \bigcirc Y) \wedge \mu W.(s \vee r \wedge \bigcirc W)$
\item[III.] $\mu X.\nu Y.(p \vee \bigcirc(X \wedge q) \vee \bigcirc(X \wedge \bigcirc Y))$
\item[IV.] $\nu Z. \bigcirc (\mu X.(\bigcirc X \vee \nu Y. (p \wedge \bigcirc Y)) \wedge \bigcirc Z)$
\item[V.] $\nu Z.(\nu X.(p\wedge \bigcirc X\vee \bigcirc\bigcirc Z)\wedge \mu Y.(q\wedge \bigcirc Y\vee r\wedge \bigcirc Z)) \wedge \nu R.(s \wedge \bigcirc\bigcirc R)$
\end{itemize}
\end{Exm}

In a PFG $G_{\phi}$ generated by our tool, when $\phi$ is satisfiable, we use a red path to denote the loop found by algorithm \emph{SCCNuSearch} which corresponds to a $\nu$-path. Therefore, any path ending with the red loop characterizes a model of $\phi$.
\begin{figure}[bp]
  \centering
  \includegraphics[scale=.28]{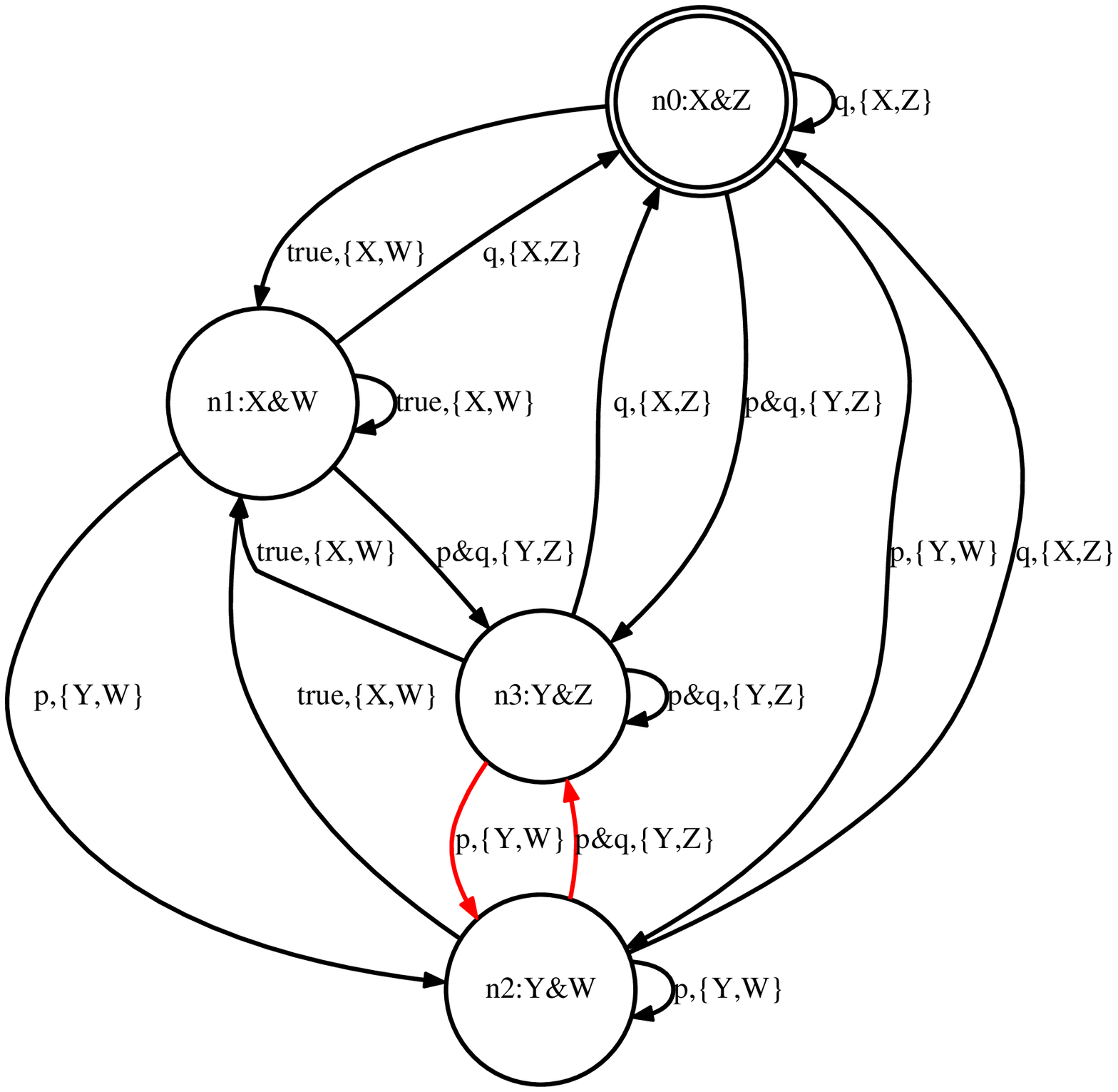}
  \caption{PFG of formula I} \label{PFGByToolExm1}
\end{figure}

\begin{figure}[tbp]
  \centering
  \includegraphics[scale=.28]{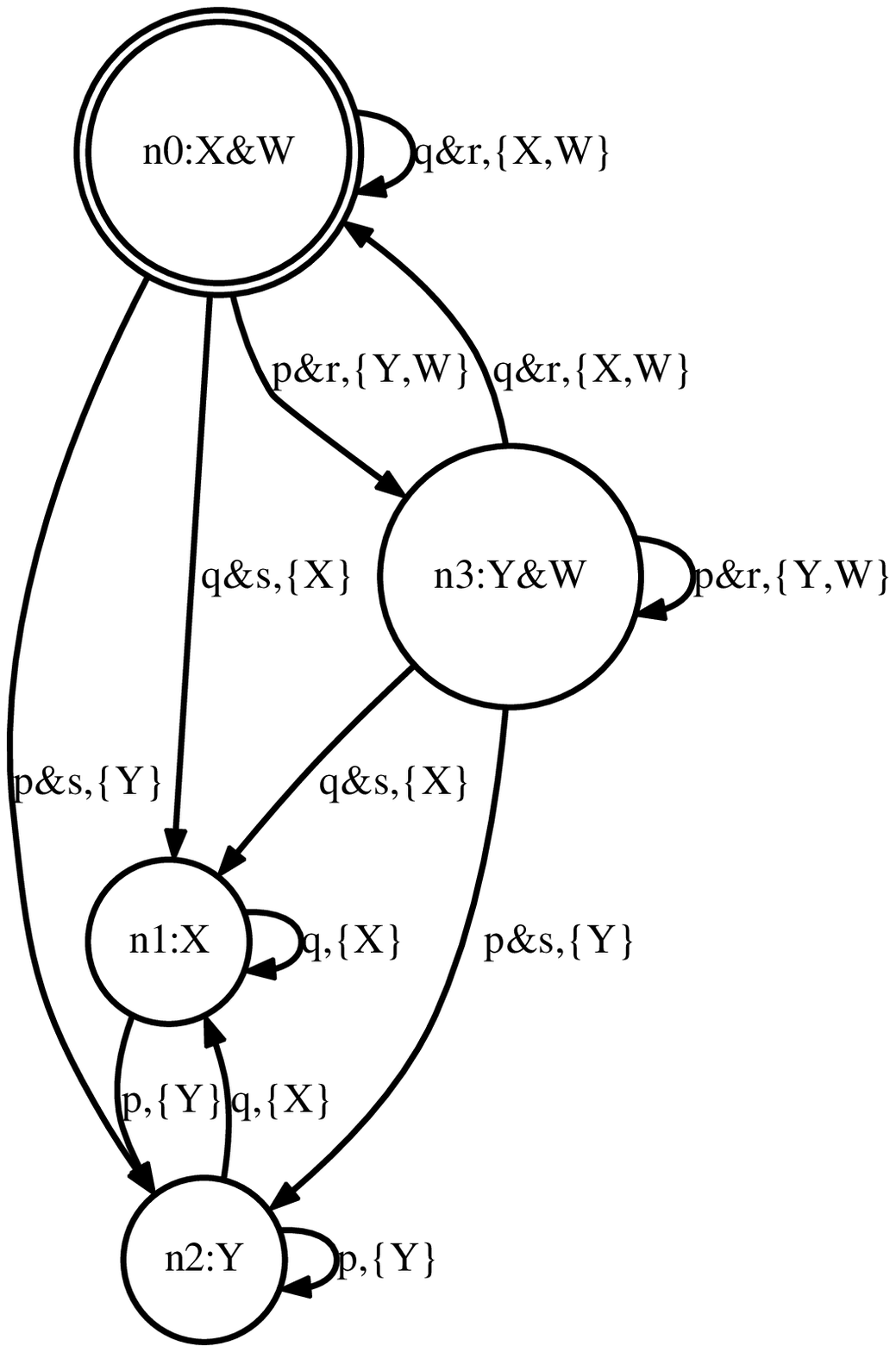}
  \caption{PFG of formula II} \label{PFGByToolExm2}
\end{figure}

\begin{figure}[tbp]
  \centering
  \includegraphics[scale=.26]{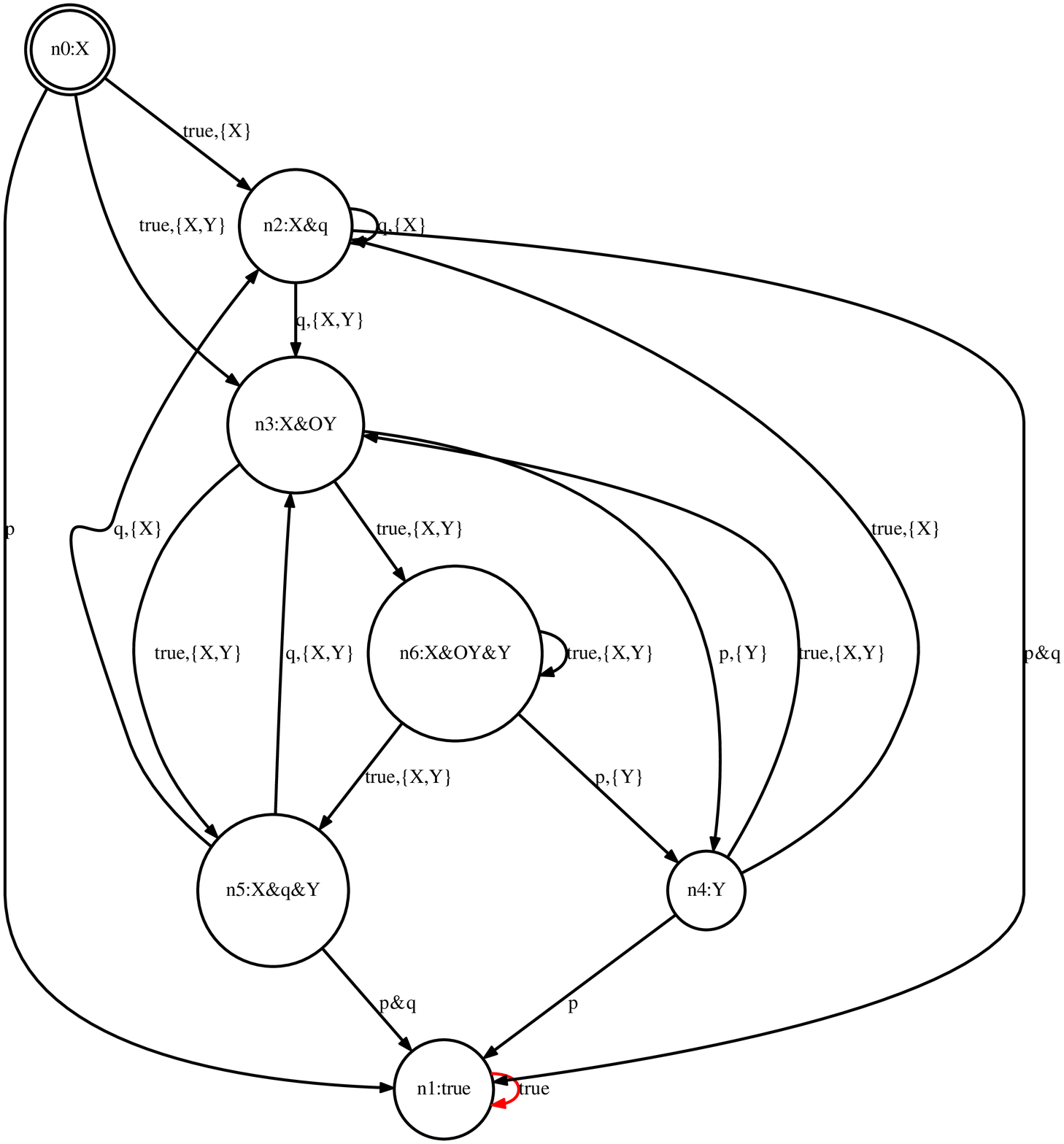}
  \caption{PFG of formula III} \label{PFGByToolExm3}
\end{figure}

\begin{figure}[tbp]
  \centering
  \includegraphics[scale=.28]{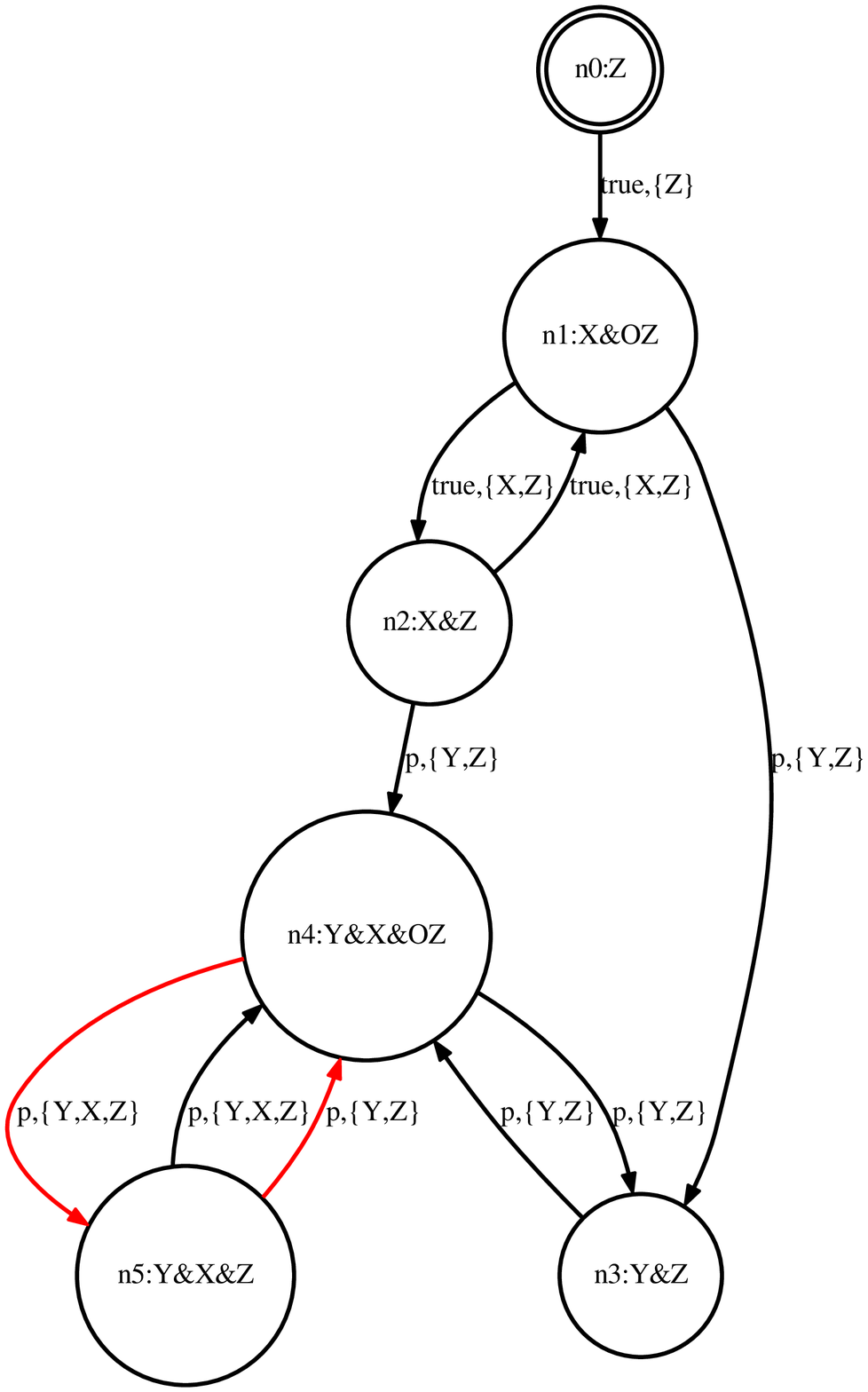}
  \caption{PFG of formula IV} \label{PFGByToolExm4}
\end{figure}

\begin{figure}[tbp]
  \centering
  \includegraphics[scale=.11]{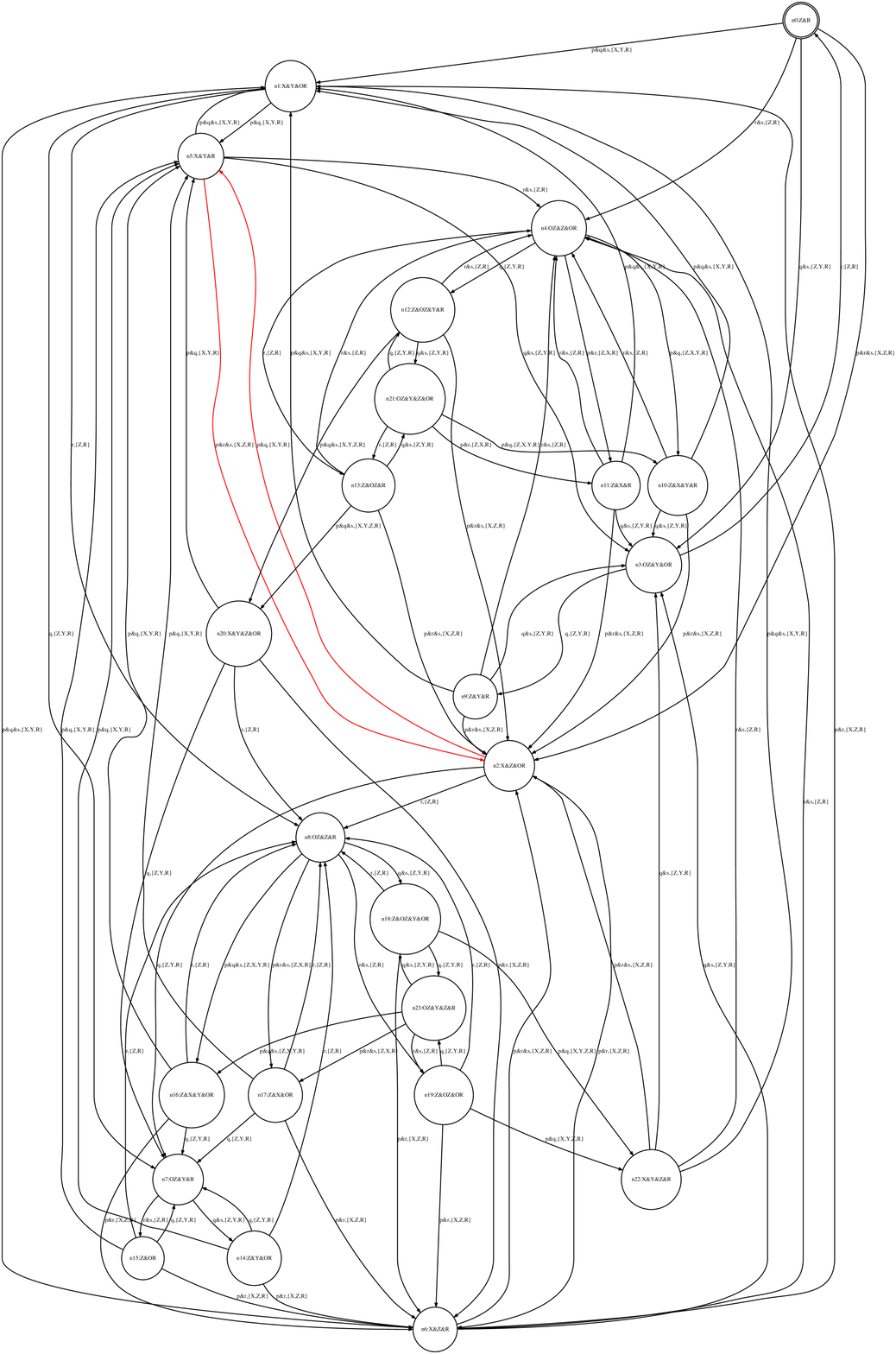}
  \caption{PFG of formula V} \label{PFGByToolExm5}
\end{figure}

As illustrated in Fig. \ref{PFGByToolExm1}, formula I is satisfiable since a $\nu$-path $n_0,p\wedge q,(n_3,p,n_2,p\wedge q)^{\omega}$ is found. The linear time structure, $\{p,q\}(\{p\}\{p,q\})^\omega$, obtained according to the $\nu$-path is indeed a model of formula I since the atomic proposition $p$ eventually always holds while $q$ always eventually holds. Similarly, from Figs. \ref{PFGByToolExm3}, \ref{PFGByToolExm4} and \ref{PFGByToolExm5} we can see that formulas III, IV and V are all satisfiable. For formula II, as depicted in Fig. \ref{PFGByToolExm2}, no red path exists in its PFG. Hence it is unsatisfiable.
\subsection{Experimental Results}
We have implemented a prototype of our PFG-based decision procedure in C++. Given a guarded formula, the prototype is able to construct its PFG and find a $\nu$-path from the PFG. To evaluate the performance of our tool, we compare it with the tool given in \cite{dhl06} which is the only tool available for the decision problems of $\nu$TL.

In \cite{dhl06}, the authors have checked the validity of the following three families of formulas: $Include_n$, $Nester_n$ and $Counter_n$ on a 1G memory PC.
\[Include_n \equiv \nu X.(\underbrace{q\wedge\bigcirc(q\wedge\bigcirc(\ldots\bigcirc(q}_{2n\; times}\wedge\bigcirc(\neg q\wedge\bigcirc X))\ldots)))\rightarrow\]
\[~~\nu Z.\mu Y.(\neg q \wedge\bigcirc Z \vee q\wedge\bigcirc(q\wedge\bigcirc Y))\]
$Nester_n \equiv \psi \vee \neg\psi$ where
\[\psi\equiv \mu X_1.\nu X_2.\mu X_3.\ldots\sigma X_n.(q_1 \vee \bigcirc(X_1 \wedge(q_2 \vee \bigcirc(X_2\wedge\ldots(q_n \vee \bigcirc X_n)\] \[\ldots))))\]
\[Counter_n \equiv \bigvee_{i=0}^{n}\neg c_i \vee \mu X.(\bigcirc X \vee (c_0 \nleftrightarrow \bigcirc\neg c_0)\vee \bigvee_{i=1}^{n}(\bigcirc c_i \nleftrightarrow c_i\wedge \neg c_{i-1}\]
\[\vee c_{i-1}\wedge(\bigcirc c_{i-1}\leftrightarrow c_i)))\]

\begin{table*}[tbp]
\caption{Experimental results}
\label{ExperimentalResults}
\centering
\begin{tabular}{|c|ccc|ccc|ccc|}

  \hline
    &   \multicolumn{3}{|c|}{$\neg Include_n$} & \multicolumn{3}{|c|}{$\neg Nester_n$} & \multicolumn{3}{|c|}{$\neg Counter_n$} \\
  \cline{2-10}
  n & Time & PFG nodes & PFG edges & Time & PFG nodes & PFG edges & Time & PFG nodes & PFG edges \\
    & (ms) & (number) & (number) & (ms) & (number) & (number) & (ms) & (number) & (number) \\
  \hline

  0 & 0 & 6 & 18 & --- & --- & --- & 0 & 2 & 2 \\
  1 & 31 & 17 & 39 & 0 & 1 & 1 & 0 & 4 & 4 \\
  2 & 63 & 28 & 64 & 31 & 10 & 30 & 16 & 8 & 8 \\
  3 & 124 & 39 & 85 & 1,185 & 73 & 386 & 156 & 16 & 16 \\
  4 & 218 & 50 & 106 & 128,559 & 601 & 4,640 & 889 & 32 & 32 \\
  5 & 328 & 61 & 127 & 18,075,924 & 5,401 & 55,419 & 5,117 & 64 & 64 \\
  \hline
\end{tabular}
\end{table*}

$Include_n$ describes the property $((aa)^{n}b)^{\omega}\subseteq ((aa)^{\ast}b)^{\omega}$, where the alphabet symbol $a$ is the label $\{q\}$ and $b$ is $\emptyset$. Note that $Include_n$ is not LTL-definable for any $n \in \mathbb{N}$. $Nester_n$ is a class of formulas with several alternating fixpoint operators. $\neg Counter_n$ formalizes an $(n+1)$-bit counter.

We equivalently check satisfiability of $\neg Include_n$, $\neg Nester_n$ and $\neg Counter_n$. The experiments are carried out on a 1.73GHz, Genuine Intel(R) CPU T2080 with 1G of memory. Table \ref{ExperimentalResults} presents the empirical measures for complexity of the PFG-based decision procedure. The columns \emph{Time} denote the running time to decide satisfiability of each formula. The columns \emph{PFG nodes} (resp. \emph{PFG edges}) represent the number of nodes (resp. edges) in the PFG of the corresponding formula. In \cite{dhl06}, the running time for checking validity of each formula is always around a few minutes. In addition, they suffer from the problem of memory overflow for formulas $Nester_4$, $Nester_5$ and $Counter_5$. However, the satisfiability in most cases can be decided in less than 1 second using our tool. In particular, it takes only about 5 seconds to decide satisfiability of $\neg Counter_5$, while the satisfiability of $\neg Nester_4$ and $\neg Nester_5$ can be decided in about 2 minutes and 5 hours, respectively. Therefore, it can be seen that our method has a better performance in practice.

\subsection{Complexity Issues}
In this section we discuss the complexity of the PFG-based decision procedure. Let $\phi$ be a closed $\nu$TL formula, $G_{\phi}=(N_{\phi},E_{\phi},n_0)$ the PFG of $\phi$, $n_v$ the number of fixpoint subformulas appearing in $\phi$. We write $|\phi|$ for the size of $\phi$, $|N_{\phi}|$ and $|E_{\phi}|$ for the number of nodes and edges in $G_{\phi}$, respectively. We can obtain, by Corollary \ref{PFGFinitenessCor}, that both $|N_{\phi}|$ and $|E_{\phi}|$ are bounded by $2^{O(|\phi|)}$. Regarding $\phi$, we have the following lemmas.

\begin{Lem}\label{PFTranRunningTimeLem}
Algorithm PFTran can be done in $2^{O(|\phi|)}$.
\end{Lem}
\textit{Proof.} First of all, it can be seen that the running time of \textit{PFTran} depends mainly on the number of recursive calls for itself as well as the running time of algorithm \textit{AND}.

The proof proceeds by induction on the structure of $\phi$.
\begin{itemize}
\item[$\bullet$] \textbf{Base case:}
\item[--] $\phi=true,false,\phi_p$, $\phi_{p}\wedge\bigcirc\varphi$ (where $\phi_{p}$ is of the form $\bigwedge_{h=1}^{n}\dot{p}_{h}$), or $\bigcirc\varphi$: the lemma holds obviously in these cases.

\item[$\bullet$] \textbf{Induction:}
\item[--] $\phi=\phi_1\vee\phi_2$: by induction hypothesis, $PFTran(\phi_1)$ and $PFTran(\phi_2)$ can be finished in $2^{O(|\phi_1|)}$ and $2^{O(|\phi_2|)}$, respectively. Further, we can see that the running time of $PFTran(\phi)$ is $2^{O(|\phi_1|)}+2^{O(|\phi_2|)}$, which is bounded by $2^{O(|\phi|)}$.

\item[--] $\phi=\phi_1\wedge\phi_2$: by induction hypothesis, $PFTran(\phi_1)$ and $PFTran(\phi_2)$ can be completed in $2^{O(|\phi_1|)}$ and $2^{O(|\phi_2|)}$, respectively. After being transformed into PF form, the number of disjuncts in $\phi_1$ (resp. $\phi_2$) is bounded by $2^{O(|\phi_1|)}$ (resp. $2^{O(|\phi_2|)}$). Hence, algorithm $AND$ can be completed in $2^{O(|\phi_1|+|\phi_2|)}$. Further, we can obtain that the overall running time of $PFTran(\phi)$ is $2^{O(|\phi_1|)}+2^{O(|\phi_2|)}+2^{O(|\phi_1|+|\phi_2|)}$, which is bounded by $2^{O(|\phi|)}$.
\item[--] $\phi=\sigma X.\varphi$: we consider only guarded formulas where each free occurrence of $X$ in $\varphi$ must be in the scope of a $\bigcirc$ operator. Regarding $X$ as an atomic proposition, $\varphi$ can be transformed into PF form by algorithm \textit{PFTran}, which can be accomplished, by induction hypothesis, in $2^{O(|\varphi|)}$. Subsequently, by substituting $\sigma X.\varphi$ for all free occurrences of $X$ in $\varphi$ (which can be done in linear time), we can obtain that $\varphi[\sigma X.\varphi/X]$ can also be transformed into PF form by algorithm \textit{PFTran} in $2^{O(|\varphi|)}$. Therefore, the running time of \textit{PFTran}$(\sigma X.\varphi)$ is bounded by $2^{O(|\phi|)}$ in this case.
\end{itemize}

It follows that algorithm \textit{PFTran} can be done in $2^{O(|\phi|)}$.
\hfill{$\Box$}

\begin{Lem}\label{PFGConstructionRunningTimeLem}
Algorithm PFGCon can be done in $2^{O(|\phi|)}$.
\end{Lem}
\textit{Proof.} The running time of \textit{PFGCon} depends mainly on three parts: (I) generating nodes and edges; (II) adding marks; (III) eliminating redundant nodes and the relative edges. In part I, since $|N_{\phi}|$ is bounded by $2^{O(|\phi|)}$, the number of iterations in Line 2 is bounded by $2^{O(|\phi|)}$. In each iteration, algorithm \textit{PFTran} is called, which can be finished in $2^{O(|\phi|)}$ according to Lemma \ref{PFTranRunningTimeLem}. Next, after the PF form transformation, we can see that the number of iterations in Line 5 of \textit{PFGCon} is bounded by $2^{O(|\phi|)}$. Hence, part I can be finished in $2^{O(|\phi|)}$. In part II, $|E_{\phi}|$ is bounded by $2^{O(|\phi|)}$. For each edge in $E_{\phi}$, we need to obtain its mark information. Algorithm \emph{AddMark} checks if a fixpoint formula, which has been unfolded by itself or a least fixpoint formula in a PF form transformation process, exists in the future part of the PF form and can be completed in $O(|\phi|)$. Therefore, part II can be completed in $2^{O(|\phi|)}$. Further, part III can apparently be finished in $2^{O(|\phi|)}$. Thus, based on the above analysis, the overall running time of \textit{PFGCon} is in $2^{O(|\phi|)}$.
\hfill{$\Box$}

\begin{Lem}\label{TarjanRunningTimeLem}
Algorithm Tarjan can be done in $2^{O(|\phi|)}$. \emph{\textbf{(\cite{tarjan73})}}
\end{Lem}
\begin{Lem}\label{isNuPathRunningTimeLem}
Algorithm isNuPath can be done in $2^{O(|\phi|)}$.
\end{Lem}
\textit{Proof.} In Line 1, the number of iterations is bounded by $2^{O(|\phi|)}$. In Line 2, the conditional statements $X \in$ \textit{Mark(e)} and $X$ \textit{is a $\mu$-variable} can be determined in $O(n_v)$ and $O(1)$, respectively. Further, the number of iterations in Line 6 (resp. Line 7) is bounded by $O(n_v)$ (resp. $2^{O(|\phi|)}$). For each $\mu$-variable $V_{\mu}$ appearing in $\phi$, we maintain a list of variables depending on $V_{\mu}$. In this way, the conditional statement in Line 8 can be decided in $O(n_v^2)$. Therefore, algorithm \textit{isNuPath} can be done in $2^{O(|\phi|)}$.
\hfill{$\Box$}
\begin{Lem}\label{SCCLoopSearchRunningTimeLem}
Algorithm SCCNuSearch can be done in $2^{O(|\phi|)}$.
\end{Lem}
\textit{Proof.} Since each edge in the input \textit{scc} is handled exactly once, the total number of recursive calls for \textit{SCCNuSearch} is bounded by $2^{O(|\phi|)}$. In Line 2, the number of iterations is also bounded by $2^{O(|\phi|)}$. Subsequently, algorithm \textit{isLoop} and \textit{isNuPath} are called as the conditional statements in Lines 6 and 8 and both of them can be determined in $2^{O(|\phi|)}$. It follows that algorithm \textit{SCCNuSearch} can be done in $2^{O(|\phi|)}$.
\hfill{$\Box$}
\begin{Thm}\label{isSatisfiableRunningTimeThm}
The decision procedure PFGSAT can be done in $2^{O(|\phi|)}$.
\end{Thm}
\textit{Proof.} This theorem is a direct consequence of Lemmas \ref{PFTranRunningTimeLem}-\ref{SCCLoopSearchRunningTimeLem}.
\hfill{$\Box$}

As far as we know, the current best time complexity for the decision problems of $\nu$TL is $2^{O(|\phi|^2\log|\phi|)}$ due to \cite{bradfield1996effective,kaivola1995simple,dhl06} and our PFG-based decision procedure noticeably improves it to $2^{O(|\phi|)}$. However, the price to pay for the improvement is exponential space.

\textit{Remarks.} Building the PFG for a given formula is similar to the process of constructing the tableau for that formula. The main difference is that marks are technically added during the PFG construction. How to add marks is guided by the condition that whether or nor the unfolding of the corresponding formulas will increase the $\mu$-signature with respect to some variable in a PF form transformation process. As a result, we can detect non-well-foundedness of the unfolding of least fixpoint formulas within a PFG. The existing decision procedures, i.e. \cite{dhl06}, usually need to construct an automaton to check non-well-foundedness. Therefore, the complexity of those methods is mainly influenced by the results from automata theory. With our method, the satisfiability of a formula can be simply decided through the PFG of the formula. Since our method is independent of the results from automata theory, we obtain a faster decision procedure. However, we have to use the information of the whole PFG when deciding satisfiability of a formula, hence our method can no longer be done in polynomial space.
\section{Model Checking Based on PFG}\label{sectionModelChecking}
In this section, we use Kripke structures as models to demonstrate how the PFG-based model checking approach for $\nu$TL is achieved.

\subsection{Kripke Structure}
Let \emph{AP} be a set of atomic propositions. A \emph{Kripke structure} \cite{Kripke1963Semantical} over \emph{AP} is defined as a quadruple $M=(S,s_{0},R,I)$ consisting of:
\begin{itemize}
\item a finite set of states $S$,
\item a designated initial state $s_{0}\in S$,
\item a transition relation $R\subseteq S\times S$ where $R$ is
\emph{total}, i.e. $\forall s\in S, \exists s'\in S$,
$(s,s')\in R$,
\item a labeling (or interpretation) function $I: S\rightarrow
2^{AP}$ defining for each state $s\in S$ the set of all atomic
propositions valid in $s$.
\end{itemize}

A \textit{path} of $M$ is an infinite sequence of states
$\rho=s_0,s_1,s_2,\ldots$, departing from the initial state $s_0$, such that for each $i\geq 0$, $(s_{i},s_{i+1}) \in R$. The word on $\rho$ is the
sequence of sets of atomic propositions
$w=I(s_0),I(s_1),I(s_2),\ldots$ which is an $\omega$-word over
alphabet $2^{AP}$.

We need to take all paths in a Kripke structure into consideration in terms of the model checking problem for $\nu$TL. Given a Kripke structure $M$ and a property $\phi$ specified by a $\nu$TL formula, we say $M\models\phi$ iff every path in $M$ satisfies $\phi$. However, when determining whether $M\models\phi$, for simplicity, we usually check whether there exists a path in $M$ satisfying $\neg\phi$: if not so, we have $M\models\phi$; otherwise, $M\not\models\phi$ and we can obtain a counterexample.

In the previous section, we have presented a decision procedure for checking satisfiability of the guarded fragment of $\nu$TL formulas based on PFG. Therefore, according to the decision procedure, we are able to formalize a PFG-based model checking approach for $\nu$TL. To do so, first, it is essential to construct the product of a Kripke structure and a PFG.

\subsection{Product Graph}
Let $M=(S,s_0,R,I)$ be a Kripke structure, $G_{\phi}=(N_{\phi},E_{\phi},n_0)$ the PFG of formula $\phi$, and $AP$ the set of atomic propositions over $M$ and $\phi$. The \textit{product} of $M$ and $G_{\phi}$ is defined as a triple $G_{M\times\phi}=(V,E,v_0)$ where:
\begin{itemize}
\item $V\subseteq S\times N_{\phi}$ is a set of nodes.
\item $E\subseteq V\times Q\times V$ is a set of edges, where $Q$ is the label of an edge between two nodes. Each $((s_i,\varphi_i),Q_i,(s_j,\varphi_j))\in E$ satisfies three conditions: (1) $(s_i,s_j)\in R$, $(\varphi_i,\varphi_e,\varphi_j)\in E_{\phi}$; (2) $Q_i\equiv\varphi_e$;
(3) $((s_i,\varphi_i),Q_i,(s_j,\varphi_j))$ has the same mark as $(\varphi_i,\varphi_e,\varphi_j)$.
\item In particular, $v_0=(s_0,n_0)$ is the root node.
\end{itemize}

In a product graph $G_{M\times\phi}$, a node is called a \textit{dead node} if it has no outgoing edge. A \textit{finite path} $\Omega=V_0,Q_0,V_1,Q_1,\ldots,V_k$ in $G_{M\times\phi}$ is a finite alternate sequence of nodes and edges starting from the root node while ending with a dead node. An \textit{infinite path} $\Omega=V_0,Q_0,V_1,Q_1,\ldots$ in $G_{M\times\phi}$ is an infinite alternate sequence of nodes and edges departing from the root node.

Given a Kripke structure $M=(S,s_{0},R,I)$ and the PFG $G_{\phi}=(N_{\phi},E_{\phi},n_0)$ of a formula $\phi$, we use algorithm \textit{PGConstruction} to construct their product.

\begin{algorithm}[h]
\caption{PGConstruction($M$, $G_{\phi}$)}\label{PGConstruction}
\begin{algorithmic}[1]
\STATE $v_0=(s_0,n_0)$, $V=\{v_0\}$, $E=\emptyset$, h[$v_0$] = 0
\WHILE{there exists $v=(s,\varphi)\in V$ and h[$v$] = 0}
\STATE h[$v$] = 1
\FOR{each $(s,s')\in R$ and $(\varphi,\varphi_e,\varphi')\in E_{\phi}$}
\IF{LabelCheck($s,\varphi_e$)}
\STATE $E=E \cup \{((s,\varphi),\varphi_e,(s',\varphi'))\}$ \quad/*the newly added edge has the same mark as edge $(\varphi,\varphi_e,\varphi')$ in $G_{\phi}$*/
\IF{$(s',\varphi') \notin V$}
\STATE $V=V \cup \{(s',\varphi')\}$
\STATE h[$(s',\varphi')$] = 0
\ENDIF
\ENDIF
\ENDFOR
\ENDWHILE
\RETURN $G_{M\times\phi}$
\end{algorithmic}
\end{algorithm}

The algorithm takes $M$ and $G_{\phi}$ as inputs and returns the product graph $G_{M\times\phi}$ in the end. The root node, $v_0$, of $G_{M\times\phi}$ is assigned to $(s_0,n_0)$. Moreover, the set of nodes $V$ and the set of edges $E$ in $G_{M\times\phi}$ are initialized to $\{v_0\}$ and empty, respectively. The algorithm repeatedly checks whether the construction could proceed on an unhandled node $v\in V$ using boolean function \textit{LabelCheck}, and then adds, if so, the corresponding nodes and edges to $V$ and $E$, respectively, until all nodes in $V$ have been handled. $h[]$ is utilized to indicate whether a node has been handled. If $h[v]=0$, $v$ needs to be further handled; otherwise, $v$ has already been handled.

\begin{algorithm}[H]
\caption{LabelCheck($s$, $\varphi_e$)}\label{CheckLabel}
\begin{algorithmic}[1]
\IF{$\varphi_e$ is $true$}
\RETURN true
\ENDIF
\FOR{each conjunct $\dot{p}$ of $\varphi_e$}
\IF{$\dot{p}$ is $p$ and $p\in I(s)$, or $\dot{p}$ is $\neg p$ and $p\notin I(s)$}
\STATE \textbf{continue}
\ELSE
\RETURN false
\ENDIF
\ENDFOR
\RETURN true
\end{algorithmic}
\end{algorithm}

Given a node $v=(s,\varphi)\in V$, we use algorithm \textit{LabelCheck} to determine if the construction could continue from a transition $(s,s')\in R$ and an edge  $(\varphi,\varphi_e,\varphi')\in E_{\phi}$. If $\varphi_e \equiv true$, the output of \textit{LabelCheck} is true; otherwise, for each conjunct $\dot{p}$ of $\varphi_{e}$, where $\dot{p}$ denotes an atomic proposition or its negation, if $\dot{p}$ is $p$ and $p\in I(s)$ (resp. $\dot{p}$ is $\neg p$ and $p\notin I(s)$), the output of \textit{LabelCheck} is true. In all other cases, the output of \textit{LabelCheck} is false. Note that when an edge $((s,\varphi),\varphi_e,(s',\varphi'))$ is added to $E$, it retains the mark of edge $(\varphi,\varphi_e,\varphi')$ in $G_{\phi}$.

Similar to the representation of a PFG, in a product graph, we also use a double circle to denote the root node and a single circle to denote each of other nodes. Each edge is denoted by a directed arc connecting two nodes. A mark is placed behind the label of an edge if it exists.

\begin{Exm}\label{PGConExm}
Constructing the product of Kripke structure $M_0$ and the PFG of formula $\phi_0$: $\bigcirc\nu X.(p\wedge \bigcirc X)$ by algorithm PGConstruction.
\end{Exm}
\begin{figure}[bp]
  \centering
  \includegraphics[scale=.6]{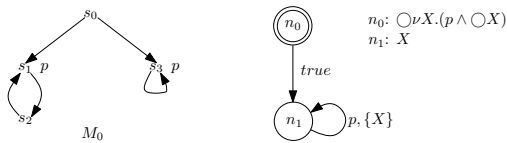}
  \caption{Kripke structure $M_0$ and the PFG of $\phi_0$} \label{PGConstructionExm1}
\end{figure}

\begin{figure}[tbp]
  \centering
  \includegraphics[scale=.6]{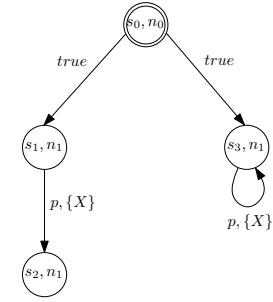}
  \caption{Product of $M_0$ and $G_{\phi_0}$} \label{PGConstructionExm2}
\end{figure}

As illustrated in Fig. \ref{PGConstructionExm2}, at the very beginning, the
root node $(s_0,n_0)$ is created and handled first by the algorithm. For the transition $(s_0,s_1)$ in $M_0$ and the edge $(n_0,true,n_1)$ in $G_{\phi_0}$, since the label of $(n_0,true,n_1)$ is $true$, the output of algorithm \textit{LabelCheck} is $true$. Therefore, node $(s_1,n_1)$ and edge $((s_0,n_0),true,(s_1,n_1))$ are created. Similarly, for the transition $(s_0,s_3)$ and the edge $(n_0,true,n_1)$, node $(s_3,n_1)$ and edge $((s_0,n_0),true,(s_3,n_1))$ are created.

Next, the algorithm deals with the node $(s_1,n_1)$. For the transition $(s_1,s_2)$ in $M_0$ and the edge $(n_1,p,n_1)$ in $G_{\phi_0}$, since $p\in I(s_1)$, the output of algorithm \textit{LabelCheck} is $true$. Therefore, node $(s_2,n_1)$ and edge $((s_1,n_1),p,(s_2,n_1))$ are created. Moreover, $((s_1,n_1),p,(s_2,n_1))$ is marked with $\{X\}$.

Subsequently, the node $(s_3,n_1)$ is handled by the algorithm. For the transition $(s_3,s_3)$ in $M_0$ and the edge $(n_1,p,n_1)$ in $G_{\phi_0}$, since $p\in I(s_3)$, the output of algorithm \textit{LabelCheck} is $true$. Thus, edge $((s_3,n_1),p,(s_3,n_1))$ is created and marked with $\{X\}$.

Further, the algorithm deals with the node $(s_2,n_1)$. For the transition $(s_2,s_1)$ in $M_0$ and the edge $(n_1,p,n_1)$ in $G_{\phi_0}$, since $p\notin I(s_2)$, the output of algorithm \textit{LabelCheck} is $false$, which indicates that the construction cannot proceed on node $(s_2,n_1)$. By now, all nodes have been handled and the whole construction process terminates.

\subsection{$\nu$-paths in Product Graph}
To formalize the PFG-based model checking approach for $\nu$TL, we apply the definition of $\nu$-paths in PFGs to the product graphs. Similarly, we concentrate only on paths ending with loops in a product graph. Given an infinite path $\Omega$ in a product graph, for convenience, we use $LES_{MC}(\Omega)$ to denote the set of edges appearing in the loop part of $\Omega$, $Mark_{MC}(e)$ the mark of edge $e$, $LMS_{MC}(\Omega)$ the set of all $\mu$-variables occurring in each $Mark_{MC}(e_i)$ where $e_i \in LES_{MC}(\Omega)$. In addition, we use $FCom(\Omega)$ to denote the sequence of the first component of each node on $\Omega$ and $SCom(\Omega)$ the alternate sequence of nodes and edges in the original PFG corresponding to the sequence of the second component of each node on $\Omega$.
\begin{Def}\label{NuPathINProductGraphDef}
Given a Kripke structure $M$ and a closed $\nu$TL formula $\phi$. An infinite path $\Omega$ in $G_{M\times\phi}$ is called a $\nu$-\emph{path} iff for each $X \in LMS_{MC}(\Omega)$, an edge $e \in LES_{MC}(\Omega)$ can be found such that $X \notin Mark_{MC}(e)$ and for any $X'$ with $X \lhd X'$, $X'\notin Mark_{MC}(e)$.
\end{Def}
\begin{Exm}\label{NuPathInPGExm}
$\nu$-paths in product graph.
\end{Exm}

\begin{figure}[bp]
  \centering
  \includegraphics[scale=.6]{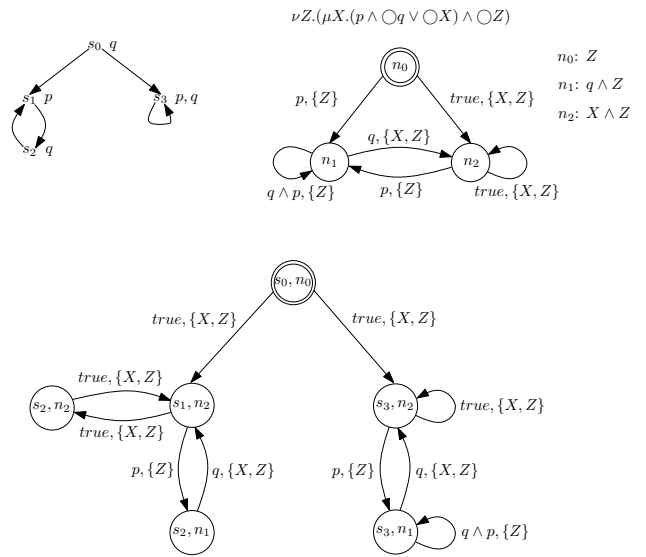}
  \caption{Examples of $\nu$-paths in product graph} \label{NuPathInPG1}
\end{figure}

\begin{itemize}
\item[1)] $\Omega_1$: $(s_0,n_0),true,(s_3,n_2),p,((s_3,n_1),q\wedge p)^\omega$. $\Omega_1$ is a $\nu$-path since $LMS_{MC}(\Omega_1)=\emptyset$.
\item[2)] $\Omega_2$: $(s_0,n_0),true,((s_1,n_2),true,(s_2,n_2),true)^\omega$. We have $LES_{MC}(\Omega_2)=\{((s_1,n_2),true,(s_2,n_2)),((s_2,n_2),true,$ $(s_1,n_2))\}$ and $LMS_{MC}(\Omega_2)=\{X\}$. For the only variable $X \in LMS_{MC}(\Omega_2)$, we cannot find an edge from $LES_{MC}(\Omega_2)$ whose mark does not contain $X$. So $\Omega_2$ is not a $\nu$-path.
\item[3)] $\Omega_3$: $(s_0,n_0),true,((s_3,n_2),p,(s_3,n_1),q)^\omega$. We have $LES_{MC}(\Omega_3)=\{((s_3,n_2),p,(s_3,n_1)),((s_3,n_1),q,(s_3,n_2))\}$ and $LMS_{MC}(\Omega_3)=\{X\}$. For the only variable $X \in LMS_{MC}(\Omega_3)$, we can find an edge $((s_3,n_2),p,(s_3,n_1))$ $\in LES_{MC}(\Omega_3)$ whose mark does not contain $X$ and any variable depending on $X$. Therefore, $\Omega_3$ is a $\nu$-path.
\end{itemize}

Regarding the notion of $\nu$-paths in a product graph, the following theorem is formalized.
\begin{Thm}\label{MCBasedOnPFGThm}
Given a Kripke structure $M$ and a closed $\nu$TL formula $\phi$. We have $M\models \phi$ iff no $\nu$-path exists in $G_{M\times \neg\phi}$.
\end{Thm}
\textit{Proof.} ($\Rightarrow$) Suppose $M\models \phi$ and there exists a $\nu$-path, $\Omega_1$, in $G_{M\times \neg\phi}$.

When $LMS_{MC}(\Omega_1)$ is empty, no infinite descending chain of $\mu$-signatures on $SCom$ $(\Omega_1)$ can be found. Thus, we have that $SCom(\Omega_1)$ characterizes a model of $\neg\phi$. That is, $FCom(\Omega_1)$ is a model of $\neg\phi$, which contradicts the premise that $M\models \phi$. Therefore, no $\nu$-paths exist in $G_{M\times \neg\phi}$ in this case.

When $LMS_{MC}(\Omega_1)$ is not empty, we can obtain that for each $Y \in LMS_{MC}(\Omega_1)$, an edge $e_1 \in LES_{MC}(\Omega_1)$ can be found such that $Y \notin Mark_{MC}(e_1)$ and there exists no $Y'\in Mark_{MC}(e_1)$ where $Y \lhd Y'$. Therefore, we can acquire the following sequence of variables relevant to $Y$ according to the sequence of marks in the loop part of $\Omega_1$:
\[Y,Y_1,Y_2,\ldots,Y_m,Y\]

Further, we can obtain the following sequence of fixpoint formulas accordingly:
\[\mu Y.\phi_{Y},\sigma Y_1.\phi_{1},\sigma Y_2.\phi_{2},\ldots,\sigma Y_m.\phi_{m},\mu Y.\phi_{Y}\]
where there must exist a formula $\sigma Y_j.\phi_{j}$ ($1\leq j \leq m$) in which $\mu Y.\phi_{Y}$ does not appear as a subformula. By the well-foundedness of $\mu$-signatures w.r.t. $Y$, we have that $SCom(\Omega_1)$ characterizes a model of $\neg\phi$. In other words, $FCom(\Omega_1)$ is a model of $\neg\phi$, which contradicts the premise that $M\models \phi$. It follows that when $M\models \phi$, there exists no $\nu$-path in $G_{M\times \neg\phi}$.

($\Leftarrow$) Let $\Omega_2$ be an arbitrary path in $G_{M\times \neg\phi}$.

When $\Omega_2$ is finite, by Theorem \ref{SatisfyThm} we know that $SCom(\Omega_2)$ does not characterize a model of $\neg\phi$. That is, any path in $M$ prefixed by $FCom(\Omega_2)$ is a model of $\phi$ in this case.

When $\Omega_2$ is infinite, there exists at least one $X \in LMS_{MC}(\Omega_2)$ such that for each edge $e_2 \in LES_{MC}(\Omega_2)$, either $X \in Mark_{MC}(e_2)$ or $X' \in Mark_{MC}(e_2)$ where $X \lhd X'$. As a result, we can obtain the following sequence of variables according to the sequence of marks in the loop part of $\Omega_2$:
\[X,X_1,X_2,\ldots,X_n,X\]
where each $X_i$ ($1\leq i \leq n$) is either $X$ itself or a variable depending on $X$.

Further, we can obtain the following sequence of fixpoint formulas accordingly:
\[\mu X.\phi_{X},\sigma X_1.\phi_{1},\sigma X_2.\phi_{2},\ldots,\sigma X_n.\phi_{n},\mu X.\phi_{X}\]
where each $\sigma X_i.\phi_{i}$ is identified by $X_i$ and $\mu X.\phi_{X}$ by $X$. Since each $X_i$ is either $X$ or a variable depending on $X$, $\mu X.\phi_{X}$ must appear as a subformula of each $\sigma X_i.\phi_{i}$. Therefore, it can be seen that the above sequence describes exactly an infinite descending chain of $\mu$-signatures w.r.t. $X$. By the well-foundedness of $\mu$-signatures, we have that $SCom(\Omega_2)$ does not characterize a model of $\neg\phi$. That is, $FCom(\Omega_2)$ is a model of $\phi$. It follows that when there exists no $\nu$-path in $G_{M\times \neg\phi}$, $M\models \phi$.
\hfill{$\Box$}

\begin{figure}[tbp]
  \centering
  \includegraphics[scale=.55]{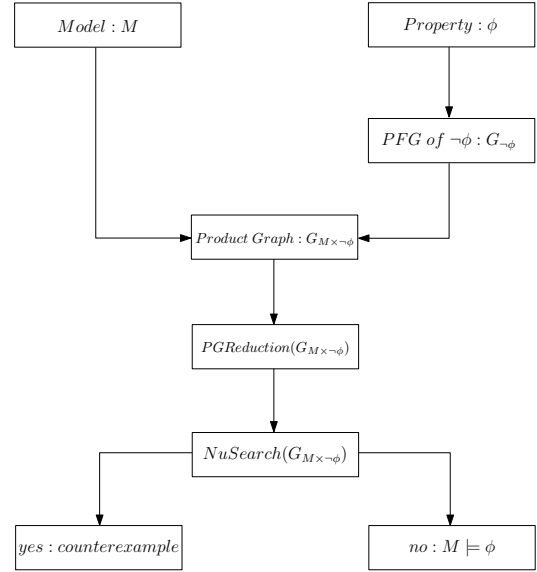}
  \caption{Model checking based on PFG} \label{MC-PFG}
\end{figure}

As a result, we reduce the model checking problem of $\nu$TL to a $\nu$-path searching problem from a product graph. According to Theorem \ref{MCBasedOnPFGThm}, we propose the PFG-based model checking process for $\nu$TL, as illustrated in Fig. \ref{MC-PFG}.

In Fig. \ref{MC-PFG}, function \emph{PGReduction} is employed to remove all dead nodes and the relative edges from the product graph $G_{M\times \neg\phi}$, while function \emph{NuSearch} is used to find a $\nu$-path in $G_{M\times \neg\phi}$. If no $\nu$-path exists in $G_{M\times \neg\phi}$, we have $M\models \phi$; otherwise, $M\not\models \phi$ and a counterexample can be obtained.

In the following we use a couple of examples to demonstrate how the PFG-based model checking approach works.
\begin{Exm}\label{MCExm1}
Checking whether Kripke structure $M_1$ satisfies property $\phi_1$: $\mu X.(p \vee \bigcirc X)\wedge \nu Y.(q\wedge\bigcirc\bigcirc Y)$.
\end{Exm}
\begin{figure}[bp]
  \centering
  \includegraphics[scale=.6]{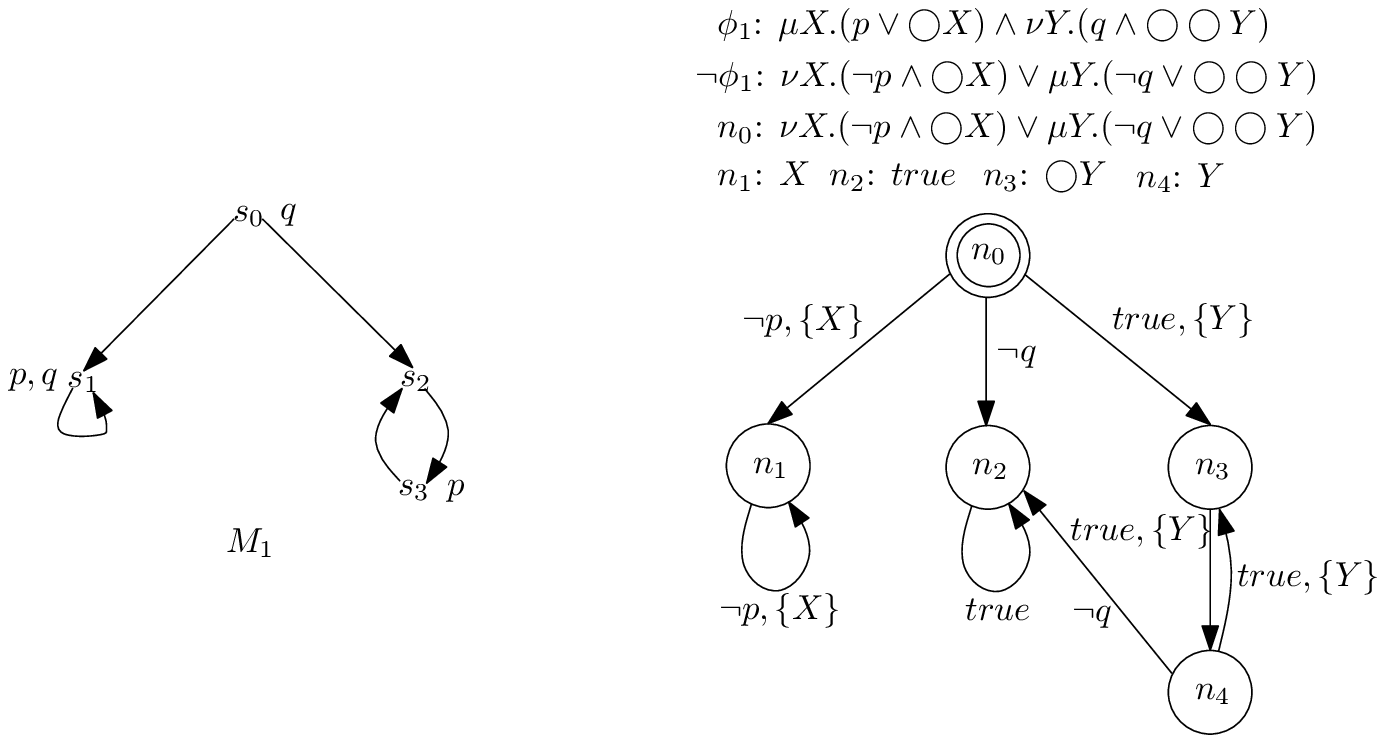}
  \caption{$M_1$ and the PFG of $\neg\phi_1$} \label{MC11}
\end{figure}

\begin{figure}[tbp]
  \centering
  \includegraphics[scale=.47]{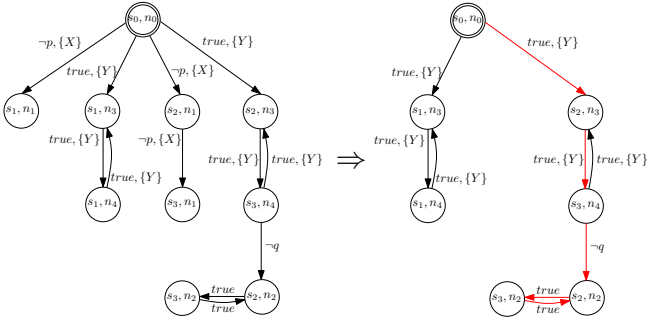}
  \caption{Product of $M_1$ and $G_{\neg\phi_1}$} \label{MC12}
\end{figure}

Here $\phi$ describes the property that $p$ finally holds and $q$ holds on every even position. The product of $M_1$ and $G_{\neg\phi_1}$ is shown in Fig. \ref{MC12}. First, we eliminate all the dead nodes and the relative edges from the product graph. After that, we try to find a $\nu$-path in the remaining graph. Since a $\nu$-path (highlighted in red) is found, we can obtain that $M_1 \not\models \phi$ and path $s_0,s_2,s_3,(s_2,s_3)^{\omega}$ in $M_1$ is a corresponding counterexample.

\begin{figure}[tbp]
  \centering
  \includegraphics[scale=.54]{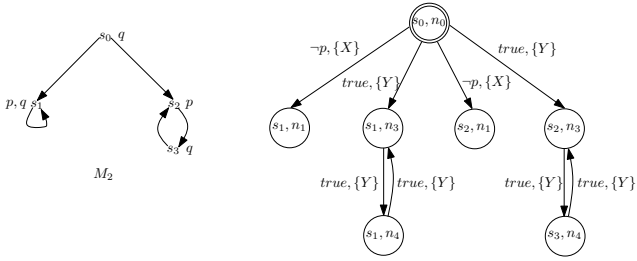}
  \caption{$M_2$ and the corresponding product graph}\label{MC13}
\end{figure}

Next, we consider Kripke structure $M_2$, as depicted in Fig. \ref{MC13}, for the same property above.
Since no $\nu$-path can be found in the product of $M_2$ and $G_{\neg\phi_1}$, we can obtain that $M_2 \models \phi$.

Further, let us consider a more complicated example.
\begin{Exm}\label{MCExm2}
Checking whether Kripke structure $M_3$ satisfies property $\phi_2$: $\nu X.\mu Y.(\bigcirc Y \vee p\wedge\bigcirc X)\vee \nu Z.(q\wedge\bigcirc\bigcirc Z)$.
\end{Exm}
\begin{figure}[tbp]
  \centering
  \includegraphics[scale=.55]{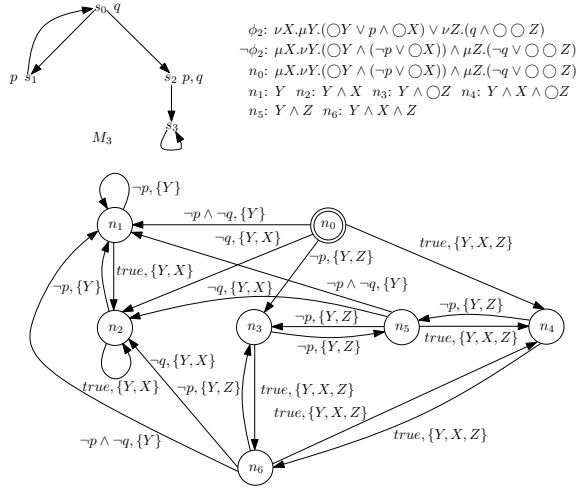}
  \caption{$M_3$ and the PFG of $\neg\phi_2$} \label{MC21}
\end{figure}

\begin{figure}[tbp]
  \centering
  \includegraphics[scale=.55]{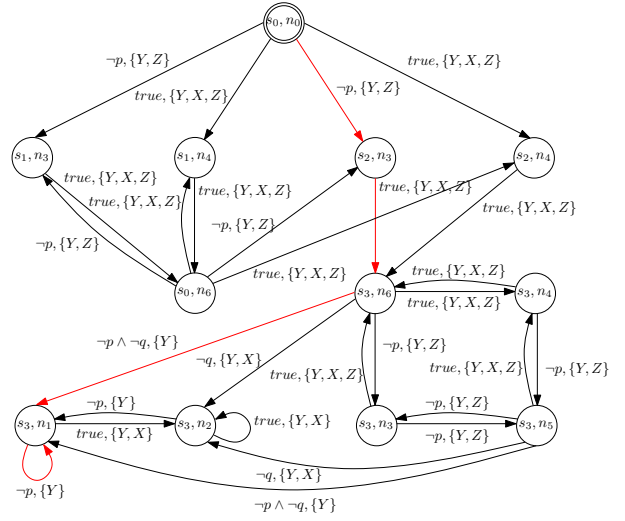}
  \caption{Product of $M_3$ and $G_{\neg\phi_2}$} \label{MC22}
\end{figure}

The product of $M_3$ and $G_{\neg\phi_2}$ is illustrated in Fig. \ref{MC22}. We can see that $M_3 \not\models \phi_2$ and path $s_0,s_2,s_3,(s_3)^{\omega}$ in $M_3$ is a counterexample.

\subsection{The Model Checking Algorithm}
In this section we present a sketch, algorithm \emph{MCPFG}, of how the PFG-based model checking approach is realized.

\begin{algorithm}[h]
\caption{MCPFG($M$, $\phi$)}
\begin{algorithmic}[1]
\STATE $G_{\neg\phi}=$ PFGCon$(\neg\phi)$
\STATE $G_{M\times\neg\phi}=$ PGConstruction($M$, $G_{\neg\phi}$)
\STATE $G_{M\times\neg\phi}=$ PGReduction($G_{M\times\neg\phi}$)
\IF{$G_{M\times\neg\phi}$ is empty}
\RETURN $M\models\phi$
\ENDIF
\STATE MCTarjan($G_{M\times\neg\phi}$, $v_0$)
\FOR {each $\mathcal{C}\in \mathcal{CS}$ }
\STATE NuSearch($v$, $\mathcal{C}$) \quad/*$v$ is an arbitrary node in $\mathcal{C}$*/
\ENDFOR
\RETURN $M\models\phi$
\end{algorithmic}
\end{algorithm}

The algorithm takes a Kripke structure $M$ and a property $\phi$, specified by a closed $\nu$TL formula, as inputs and eventually returns the result whether $M\models\phi$. To do so, the PFG, $G_{\neg\phi}$, of $\neg\phi$ is constructed first by algorithm \textit{PFGCon}. Next, the product of $M$ and $G_{\neg\phi}$ is constructed by algorithm \textit{PGConstruction} and then reduced by algorithm \textit{PGReduction} (as shown in Algorithm \ref{PGReduction}). If the reduced product graph $G_{M\times\neg\phi}$ is empty, we have $M\models\phi$ since no $\nu$-path can be found in $G_{M\times\neg\phi}$; otherwise, the algorithm will try to find a $\nu$-path in $G_{M\times\neg\phi}$. Further, algorithm \textit{MCTarjan} is employed to compute all SCCs in $G_{M\times\neg\phi}$. Finally, the algorithm checks whether there exists a loop in some SCC which corresponds to a $\nu$-path by algorithm \textit{NuSearch}: if so, \textit{NuSearch} will return that $M\not\models\phi$ and a corresponding counterexample can be obtained; otherwise, $M\models\phi$.

\begin{algorithm}[htbp]
\caption{PGReduction($G_{M\times\neg\phi}$)}\label{PGReduction}
\begin{algorithmic}[1]
\FORALL{$v \in V$ with no outgoing edge}
\STATE $V=V\setminus \{v\}$ \quad/*eliminating dead nodes and the relative edges*/
\STATE $E=E \setminus \bigcup_{i}\{(v_i,Q_i,v)\}$
\ENDFOR
\RETURN $G_{M\times\neg\phi}$
\end{algorithmic}
\end{algorithm}

Note that algorithm \textit{NuSearch} uses boolean function \textit{isNu} to determine whether a sequence of edges in $G_{M\times\neg\phi}$ corresponds to a $\nu$-path. Algorithms \textit{MCTarjan}, \textit{NuSearch} and \textit{isNu} are similar to algorithms \textit{Tarjan}, \textit{SCCNuSearch} and \textit{isNuPath}, respectively, except that we consider product graphs here instead of PFGs.

\subsection{Complexity Issues}
In this section we discuss the complexity of the PFG-based model checking approach for $\nu$TL. Let $M=(S,s_0,R,I)$ be a Kripke structure, $\phi$ a property specified by a closed $\nu$TL formula, $G_{\neg\phi}=\{N_{\neg\phi},E_{\neg\phi},n_0\}$ the PFG of $\neg\phi$, and $N_V$ (resp. $N_p$) the number of fixpoint subformulas (resp. atomic propositions) appearing in $\phi$. We write $|S|$ (resp. $|R|$) for the number of states (resp. transitions) in $M$, $|\phi|$ for the size of $\phi$, and $|N_{\neg\phi}|$ (resp. $|E_{\neg\phi}|$) for the number of nodes (resp. edges) in $G_{\neg\phi}$, respectively. Note that $|S|$ is no larger than $|R|$ since $R$ is total. According to Corollary \ref{PFGFinitenessCor}, it is easy to see that both $|N_{\neg\phi}|$ and $|E_{\neg\phi}|$ are bounded by $2^{O(|\phi|)}$. Therefore, the number of nodes (resp. edges) in $G_{M\times \neg\phi}$ is bounded by $O(|S|)\cdot 2^{O(|\phi|)}$ (resp. $O(|R|)\cdot 2^{O(|\phi|)}$). Regarding $M$ and $\phi$, we have the following lemmas.

\begin{Lem}\label{MCPGConstructionRunningTimeLem}
Algorithm PGConstruction can be done in $O(|S|^2\cdot |R|)\cdot 2^{O(|\phi|)}$.
\end{Lem}
\textit{Proof.} For each unhandled node $v$ in $G_{M\times \neg\phi}$, the algorithm checks whether new nodes and edges can be generated due to $v$. Therefore, the number of iterations of the \textit{while loop} is bounded by $O(|S|)\cdot 2^{O(|\phi|)}$. Next, it is easy to see that the number of iterations of the \textit{for loop} is bounded by $O(|R|)\cdot 2^{O(|\phi|)}$. In each iteration of the \textit{for loop}, function \textit{LabelCheck} is called to decide whether the construction could proceed on the node currently being handled, which can apparently be finished in $O(N_p^2)$. Further, the conditional statement in Line 7 of \textit{PGConstruction} can be determined in $O(|S|)\cdot 2^{O(|\phi|)}$. Therefore, algorithm \textit{PGConstruction} can be done in $O(|S|^2\cdot |R|)\cdot 2^{O(|\phi|)}$.
\hfill{$\Box$}

In addition, the following lemma is straightforward.
\begin{Lem}\label{MCPGReductionRunningTimeLem}
Algorithm PGReduction can be done in $O(|S|\cdot |R|)\cdot 2^{O(|\phi|)}$.
\end{Lem}

\begin{Lem}\label{MCTarjanRunningTimeLem}
Algorithm MCTarjan can be done in $O(|S|+|R|)\cdot 2^{O(|\phi|)}$, namely $O(|R|)\cdot 2^{O(|\phi|)}$. \emph{\textbf{(\cite{tarjan73})}}
\end{Lem}

\begin{Lem}\label{MCisNuRunningTimeLem}
Algorithm isNu can be done in $O(|R|)\cdot 2^{O(|\phi|)}$.
\end{Lem}
\textit{Proof.} Let $ES_{MC}$ be the input to \textit{isNu} where $ES_{MC}$ is a sequence of edges in $G_{M\times \neg\phi}$. The algorithm first obtains the set of $\mu$-variables, $MU$, occurring in each $Mark_{MC}(e)$ where $e\in ES_{MC}$, which can be completed in $O(|R|)\cdot 2^{O(|\phi|)}$. Subsequently, for each $V\in MU$, the algorithm tries to find an edge $e'\in ES_{MC}$ satisfying the following condition: $V\notin Mark_{MC}(e')$ and $V'\notin Mark_{MC}(e')$ where $V\lhd V'$. By maintaining, for each $\mu$-variable $Y$ appearing in $\neg\phi$, a list of variables depending on $Y$, it is not hard to see that this condition can be decided in $O(N_V^2)$. Therefore, the running time of this part is in $O(|R|)\cdot 2^{O(|\phi|)}$. It follows that algorithm \textit{isNu} can be done in $O(|R|)\cdot 2^{O(|\phi|)}$.
\hfill{$\Box$}
\begin{Lem}\label{MCNuSearchRunningTimeLem}
Algorithm NuSearch can be done in $O(|R|^3)\cdot 2^{O(|\phi|)}$.
\end{Lem}
\textit{Proof.} Let $v$ and $\mathcal{C}$ be the inputs to \textit{NuSearch} where $\mathcal{C}$ is an SCC in $G_{M\times \neg\phi}$ and $v$ is a node in $\mathcal{C}$. The algorithm calls itself recursively to build a path starting from $v$ which is likely to correspond to a $\nu$-path in $G_{M\times \neg\phi}$. Since each edge in $\mathcal{C}$ is handled exactly once, the total number of recursive calls for \textit{NuSearch} is bounded by $O(|R|)\cdot 2^{O(|\phi|)}$. Further, for each unvisited edge $e$ in $\mathcal{C}$ which takes the input node as its source node, the algorithm adds $e$ to a vector $EV$ and then checks whether there exists a loop in $EV$. It is obvious that checking the existence of a loop can be completed in $O(|S|)\cdot 2^{O(|\phi|)}$. If such a loop does exist, the algorithm calls \textit{isNu} to determine whether it corresponds to a $\nu$-path, which can be accomplished in $O(|R|)\cdot 2^{O(|\phi|)}$ by Lemma \ref{MCisNuRunningTimeLem}; otherwise, a recursive call is made. Therefore, this part can be finished in $O(|S|\cdot |R|+ |R|^2)\cdot 2^{O(|\phi|)}$, namely $O(|R|^2)\cdot 2^{O(|\phi|)}$. It follows that algorithm \textit{NuSearch} can be done in $O(|R|^3)\cdot 2^{O(|\phi|)}$.
\hfill{$\Box$}

\begin{Thm}\label{MCPFGRunningTimeThm}
The model checking algorithm MCPFG can be done in $O(|S|\cdot |R|^3)\cdot 2^{O(|\phi|)}$.
\end{Thm}
\textit{Proof.} Since the total number of SCCs in $G_{M\times \neg\phi}$ is bounded by $O(|S|)\cdot 2^{O(|\phi|)}$, by Lemmas \ref{PFGConstructionRunningTimeLem} and \ref{MCPGConstructionRunningTimeLem}-\ref{MCNuSearchRunningTimeLem} we can obtain that algorithm \textit{MCPFG} can be done in $O(|S|\cdot |R|^3)\cdot 2^{O(|\phi|)}$.
\hfill{$\Box$}
\section{Related Work}\label{sectionRelatedWork}
The major milestone of the decision problems for modal $\mu$-calculus is made by Streett and Emerson \cite{streett1989automata} who introduce the notions of choice functions, signatures and well-founded pre-models, and apply automata theory to check satisfiability. They show that a formula is satisfiable iff it has a well-founded pre-model. Two automata, $\mathcal{A}$ and $\mathcal{B}$, are used in their decision procedure. $\mathcal{A}$ checks the consistence of pre-models while $\mathcal{B}$ detects non-well-foundedness of least fixpoints. The decision procedure is finally achieved by doing an emptiness test for the product automaton $\mathcal{A}\times \overline{\mathcal{B}}$. Related methods \cite{EJS93,KVW00} translate a formula into an equivalent alternating tree automaton and then check for emptiness.

In \cite{vardi1988temporal}, Vardi first adapts Streett and Emerson's method to $\nu$TL with past operators. In his work, two-way automata are used to deal with the past operators and an algorithm running in $2^{O(|\phi|^4)}$ is obtained eventually. In \cite{Banieqbal1987}, Banieqbal and Barringer show that if a formula has a model, then it is able to generate a good Hintikka structure which can be further transformed into a good path searching problem from a graph. Their algorithm is equivalent in time complexity to Vardi's but runs in exponential space.

Stirling and Walker \cite{stirling1990ccs} first present a tableau characterisation for $\nu$TL's decision problems. However, they do not give any complexity analysis due to the complicated success conditions. Later, Bradfield, Esparza and Mader \cite{bradfield1996effective} improve the system of Stirling and Walker by simplifying the success conditions for a tableau. In their system a successful terminal is determined by the path leading to it, whereas Stirling and Walker's method requires the examination of a potentially infinite number of paths extending over the whole tableau. Using standard results from complexity theory, they obtain an algorithm running in $2^{O(|\phi|^2\log|\phi|)}$. Moreover, their system uses a couple of similar notions in \cite{kaivola1995simple} but gets rid of the use of recurrence points which will lead to a significant increase in the number of possible tableaux for a given root. A tableau system for modal $\mu$-calculus which does not rely on automata theory is given in \cite{jungteerapanich2009tableau} where the notion of \emph{names} is used to keep track of the unfolding of fixpoint variables. In \cite{friedmann2013deciding}, a tableau calculus for deciding satisfiability of arbitrary formulas is presented based on a new unfolding rule for greatest fixpoint formulas which allows unguarded formulas to be handled without an explicit transformation into guarded form.

In \cite{dhl06}, Dax, Hofmann and Lange present a simple proof system for $\nu$TL. In the system, a sequent is a subset of the closure of a formula $\phi$ and semantically stands for the disjunction of elements in the closure. A pre-proof for $\phi$ is a possibly infinite tree whose nodes are labeled with sequents, whose root is labeled with $\vdash \phi$, which is built by the corresponding proof rules. For an infinite branch $\pi$ in a pre-proof for $\phi$, they define the notion of $\nu$-$threads$ contained in $\pi$. Moreover, they show that a proof for $\phi$ is a pre-proof where every finite branch ends with a \textit{true} sequent, and every infinite branch contains a $\nu$-thread. To check if there exists a proof for $\phi$ (or validity of $\phi$), they use a nondeterministic parity automaton $\mathcal{A}_{\phi}$ to accept exactly the branches which contain a $\nu$-thread, and a deterministic B\"{u}chi automaton $\mathcal{B}_{\phi}$ to accept all the words which form a branch in a pre-proof for $\phi$. Further, they prove that for any $\nu$TL formula $\phi$, $L(\mathcal{B}_{\phi})\subseteq L(\mathcal{A}_{\phi})$ iff $\vdash\phi$. Therefore, it suffices to check the language $L(\mathcal{B}_{\phi})\cap \overline{L(\mathcal{A}_{\phi})}$ for non-emptiness, which can be done in PSPACE \cite{sistla1985complexity}. Depending on which complementation procedure is used they obtain an algorithm running in $2^{O(|\phi|^2\log|\phi|)}$ and implement it in OCAML.

In our method, given a formula $\phi$, by repeating PF form transformations, we build the PFG $G_{\phi}$ describing the possible models of $\phi$. The process of constructing $G_{\phi}$ guarantees that each node in $G_{\phi}$ corresponds to a consistent subset of $CL(\phi)$. Meanwhile, during the construction, we technically add marks to $G_{\phi}$ which will be used to trace the infinite unfolding problem, i.e. non-well-foundedness, for least fixpoint formulas. Based on those marks, we define the notion of $\nu$-paths and show that $\phi$ is satisfiable iff a $\nu$-path is contained in $G_{\phi}$. Therefore, we no longer need an automaton to detect non-well-foundedness of least fixpoint formulas. Since our method avoids the use of any result from automata or complexity theory, we obtain a faster procedure. However, when checking satisfiability of a formula, we need to store the whole PFG of the formula. Thus, our method runs in exponential space.

\section{Conclusion}\label{sectionConclusion}
In this paper, we have proved that every closed $\nu$TL formula can be
transformed into a PF form whose future part is the
conjunction of elements in the closure of a given formula. We have presented an algorithm
for constructing PFG and a decision procedure for checking
satisfiability of the guarded fragment of $\nu$TL formulas based on PFG. Also, we have implemented the decision procedure in C++. Experimental results show that our procedure performs better than the one given in \cite{dhl06}. Moreover, we have proposed a model checking approach for $\nu$TL based on PFG. Compared with the existing methods for checking satisfiability of $\nu$TL formulas, our decision procedure has several advantages: (1) it does not rely on automata theory by considering PFGs; (2) it is more efficient in time and practical meanwhile; (3) it gives good insight into why and how a given formula is satisfiable through its PFG; (4) it visually reflects that why a path is a counterexample through the corresponding product graph when a Kripke structure violates a property.

In the near future, we intend to improve the performance of our decision procedure by technically choosing outgoing edges when trying to find a $\nu$-path. We will also develop a practical PFG-based model checker for $\nu$TL and do some further case studies for more complex systems and properties.

\ifCLASSOPTIONcaptionsoff
  \newpage
\fi

\bibliographystyle{IEEEtran}
\bibliography{IEEEabrv,liuyao}

\begin{thebibliography}{10}
\providecommand{\url}[1]{#1}
\csname url@samestyle\endcsname
\providecommand{\newblock}{\relax}
\providecommand{\bibinfo}[2]{#2}
\providecommand{\BIBentrySTDinterwordspacing}{\spaceskip=0pt\relax}
\providecommand{\BIBentryALTinterwordstretchfactor}{4}
\providecommand{\BIBentryALTinterwordspacing}{\spaceskip=\fontdimen2\font plus
\BIBentryALTinterwordstretchfactor\fontdimen3\font minus
  \fontdimen4\font\relax}
\providecommand{\BIBforeignlanguage}[2]{{%
\expandafter\ifx\csname l@#1\endcsname\relax
\typeout{** WARNING: IEEEtran.bst: No hyphenation pattern has been}%
\typeout{** loaded for the language `#1'. Using the pattern for}%
\typeout{** the default language instead.}%
\else
\language=\csname l@#1\endcsname
\fi
#2}}
\providecommand{\BIBdecl}{\relax}
\BIBdecl

\bibitem{barringer1986really}
H.~Barringer, R.~Kuiper, and A.~Pnueli, ``A really abstract concurrent model
  and its temporal logic,'' in \emph{Conference Record of the 13th Annual ACM
  Symposium on Principles of Programming Languages}.\hskip 1em plus 0.5em minus
  0.4em\relax ACM, 1986, pp. 173--183.

\bibitem{Koz83}
D.~Kozen, ``\BIBforeignlanguage{English}{Results on the propositional
  $\mu$-calculus},'' \emph{\BIBforeignlanguage{English}{Theoretical Computer
  Science}}, vol.~27, no.~3, pp. 333--354, 1983.

\bibitem{Pnu77}
A.~Pnueli, ``\BIBforeignlanguage{English}{The temporal logic of programs},'' in
  \emph{\BIBforeignlanguage{English}{Proceedings of the 18th Annual Symposium
  on Foundations of Computer Science}}.\hskip 1em plus 0.5em minus 0.4em\relax
  IEEE, 1977, pp. 46--57.

\bibitem{EC80}
E.~A. Emerson and E.~M. Clarke, ``Characterizing correctness properties of
  parallel programs using fixpoints,'' in \emph{Proceedings of the 7th
  International Colloquium on Automata, Languages and Programming}, vol. 85 of
  LNCS.\hskip 1em plus 0.5em minus 0.4em\relax Springer, 1980, pp. 169--181.

\bibitem{barringer1985compositional}
H.~Barringer, R.~Kuiper, and A.~Pnueli, ``A compositional temporal approach to
  a csp-like language,'' \emph{Formal Models of Programming}, pp. 207--227,
  1985.

\bibitem{clarke1986automatic}
E.~M. Clarke, E.~A. Emerson, and A.~P. Sistla, ``Automatic verification of
  finite-state concurrent systems using temporal logic specifications,''
  \emph{ACM Transactions on Programming Languages and Systems}, vol.~8, no.~2,
  pp. 244--263, 1986.

\bibitem{vardi1988temporal}
M.~Y. Vardi, ``A temporal fixpoint calculus,'' in \emph{Conference Record of
  the 15th Annual ACM Symposium on Principles of Programming Languages}.\hskip
  1em plus 0.5em minus 0.4em\relax ACM, 1988, pp. 250--259.

\bibitem{streett1989automata}
R.~S. Streett and E.~A. Emerson, ``An automata theoretic decision procedure for
  the propositional mu-calculus,'' \emph{Information and Computation}, vol.~81,
  no.~3, pp. 249--264, 1989.

\bibitem{EJS93}
E.~A. Emerson, C.~S. Jutla, and A.~P. Sistla, ``On model-checking for fragments
  of $\mu$-calculus,'' in \emph{Proceedings of the 5th International Conference
  on Computer Aided Verification}, vol. 697 of LNCS.\hskip 1em plus 0.5em minus
  0.4em\relax Springer, 1993, pp. 385--396.

\bibitem{KVW00}
O.~Kupferman, M.~Y. Vardi, and P.~Wolper, ``An automata-theoretic approach to
  branching-time model checking,'' \emph{Journal of the Association for
  Computing Machinery}, vol.~47, no.~2, pp. 312--360, 2000.

\bibitem{Banieqbal1987}
B.~Banieqbal and H.~Barringer, ``Temporal logic with fixed points,'' in
  \emph{Proceedings of the Collection on Temporal Logic in Specification}, vol.
  398 of LNCS.\hskip 1em plus 0.5em minus 0.4em\relax Springer, 1989, pp.
  62--74.

\bibitem{stirling1990ccs}
C.~Stirling and D.~Walker, ``Ccs, liveness, and local model checking in the
  linear time mu-calculus,'' in \emph{Proceedings of the 1st International
  Workshop on Automatic Verification Methods for Finite State Systems}, vol.
  407 of LNCS.\hskip 1em plus 0.5em minus 0.4em\relax Springer, 1990, pp.
  166--178.

\bibitem{bradfield1996effective}
J.~Bradfield, J.~Esparza, and A.~Mader, ``An effective tableau system for the
  linear time $\mu$-calculus,'' in \emph{Proceedings of the 23rd International
  Colloquium on Automata, Languages and Programming}, vol. 1099 of LNCS.\hskip
  1em plus 0.5em minus 0.4em\relax Springer, 1996, pp. 98--109.

\bibitem{kaivola1995simple}
R.~Kaivola, ``A simple decision method for the linear time mu-calculus,'' in
  \emph{Proceedings of the International Workshop on Structures in Concurrency
  Theory}.\hskip 1em plus 0.5em minus 0.4em\relax Springer, 1995, pp. 190--204.

\bibitem{dhl06}
C.~Dax, M.~Hofmann, and M.~Lange, ``A proof system for the linear time
  $\mu$-calculus,'' in \emph{Proceedings of the 26th International Conference
  on Foundations of Software Technology and Theoretical Computer Science}, vol.
  4337 of LNCS.\hskip 1em plus 0.5em minus 0.4em\relax Springer, 2006, pp.
  274--285.

\bibitem{PFForm14}
Y.~Liu, Z.~Duan, C.~Tian, and B.~Liu, ``Present-future form of linear time
  $\mu$-calculus,'' in \emph{Proceedings of the 3rd International Workshop on
  SOFL+MSVL}, vol. 8332 of LNCS.\hskip 1em plus 0.5em minus 0.4em\relax
  Springer, 2013, pp. 76--85.

\bibitem{duan1996extended}
Z.~Duan, ``An extended interval temporal logic and a framing technique for
  temporal logic programming,'' Ph.D. dissertation, University of Newcastle
  Upon Tyne, 1996.

\bibitem{duan2006book}
Z.~Duan, \emph{Temporal Logic and Temporal Logic Programming}.\hskip 1em plus
  0.5em minus 0.4em\relax China: Science Press, 2006.

\bibitem{DT2007Decidability}
Z.~Duan and C.~Tian, ``Decidability of propositional projection temporal logic
  with infinite models,'' in \emph{Proceedings of the 4th International
  Conference on Theory and Applications of Models of Computation}, vol. 4484 of
  LNCS.\hskip 1em plus 0.5em minus 0.4em\relax Springer, 2007, pp. 521--532.

\bibitem{DT2010ImprovedDecision}
Z.~Duan and C.~Tian, ``An improved decision procedure for propositional
  projection temporal logic,'' in \emph{Proceedings of the 12th International
  Conference on Formal Engineering Methods}, vol. 6447 of LNCS.\hskip 1em plus
  0.5em minus 0.4em\relax Springer, 2010, pp. 90--105.

\bibitem{DTpracticaldecision14}
Z.~Duan and C.~Tian, ``A practical decision procedure for propositional
  projection temporal logic with infinite models,'' \emph{Theoretical Computer
  Science}, vol. 554, no.~0, pp. 169--190, 2014.

\bibitem{stirling1991modal}
C.~Stirling, \emph{Modal and temporal logics}.\hskip 1em plus 0.5em minus
  0.4em\relax LFCS, Department of Computer Science, University of Edinburgh,
  1991.

\bibitem{walukiewicz2000completeness}
I.~Walukiewicz, ``Completeness of kozen's axiomatisation of the propositional
  $\mu$-calculus,'' \emph{Information and Computation}, vol. 157, no.~1, pp.
  142--182, 2000.

\bibitem{bruse2015guarded}
F.~Bruse, O.~Friedmann, and M.~Lange, ``On guarded transformation in the modal
  $\mu$-calculus,'' \emph{Logic Journal of the IGPL}, vol.~23, no.~2, pp.
  194--216, 2015.

\bibitem{tarski1955lattice}
A.~Tarski, ``A lattice-theoretical fixpoint theorem and its applications,''
  \emph{Pacific Journal of Mathematics}, vol.~5, no.~2, pp. 285--309, 1955.

\bibitem{FL79}
M.~J. Fischer and R.~E. Ladner, ``Propositional dynamic logic of regular
  programs,'' \emph{Journal of Computer and System Sciences}, vol.~18, no.~2,
  pp. 194--211, 1979.

\bibitem{tarjan73}
R.~Tarjan, ``\BIBforeignlanguage{English}{Depth-first search and linear graph
  algorithms},'' \emph{\BIBforeignlanguage{English}{SIAM Journal on
  Computing}}, vol.~1, no.~2, pp. 146--160, 1973.

\bibitem{Kripke1963Semantical}
S.~A. Kripke, ``Semantical analysis of modal logic i: Normal propositional
  calculi,'' \emph{Zeitschrift Fur Mathematische Logik Und Grundlagen Der
  Mathematik}, vol.~9, no. 5-6, pp. 67--96, 1963.

\bibitem{jungteerapanich2009tableau}
N.~Jungteerapanich, ``A tableau system for the modal $\mu$-calculus,'' in
  \emph{Proceedings of the 18th International Conference on Automated Reasoning
  with Analytic Tableaux and Related Methods}, vol. 5607 of LNCS.\hskip 1em
  plus 0.5em minus 0.4em\relax Springer, 2009, pp. 220--234.

\bibitem{friedmann2013deciding}
O.~Friedmann and M.~Lange, ``Deciding the unguarded modal $\mu$-calculus,''
  \emph{Journal of Applied Non-Classical Logics}, vol.~23, no.~4, pp. 353--371,
  2013.

\bibitem{sistla1985complexity}
A.~P. Sistla and E.~M. Clarke, ``The complexity of propositional linear
  temporal logics,'' \emph{Journal of the Association for Computing Machinery},
  vol.~32, no.~3, pp. 733--749, 1985.

\end{thebibliography}


\begin{thebibliography}{1}

\bibitem{IEEEhowto:kopka}
H.~Kopka and P.~W. Daly, \emph{A Guide to \LaTeX}, 3rd~ed.\hskip 1em plus
  0.5em minus 0.4em\relax Harlow, England: Addison-Wesley, 1999.

\end{thebibliography}
\end{document}